\documentstyle{mn}

\newif\ifAMStwofonts
\AMStwofontstrue

\input{epsf}

\def\wfe{W_{\rm Fe}}
\def\al{\alpha}
\def\as{\alpha_{\rm s}}

\def\g{$\gamma$}
\def\nh{N_{\rm H}}

\def\af{A_{\rm Fe}}
\def\taut{\tau_{\rm T}}
\def\exosat{{\it EXOSAT}}
\def\rosat{{\it ROSAT}}
\def\asca{{\it ASCA}}
\def\ginga{{\it Ginga}}
\def\heao{{\it HEAO-1}}
\def\ec{E_{\rm c}}
\def\eb{E_{\rm b}}
\def\fek{Fe K}
\def\efe{E_{\rm Fe}}
\def\sfe{\sigma_{\rm Fe}}
\def\ife{I_{\rm Fe}}
\def\cm_2{cm$^{-2}$}
\def\cd{\chi^2/{\rm dof}}
\def\cnu{\chi^2_\nu}

\ifoldfss
  \ifCUPmtlplainloaded \else
    \NewTextAlphabet{textbfit} {cmbxti10} {}
    \NewTextAlphabet{textbfss} {cmssbx10} {}
    \NewMathAlphabet{mathbfit} {cmbxti10} {} 
    \NewMathAlphabet{mathbfss} {cmssbx10} {} 
  \fi
  \ifAMStwofonts
    \ifCUPmtlplainloaded \else
      \NewSymbolFont{upmath} {eurm10}
      \NewSymbolFont{AMSa} {msam10}
      \NewMathSymbol{\upi}     {0}{upmath}{19}
      \NewMathSymbol{\umu}     {0}{upmath}{16}
      \NewMathSymbol{\upartial}{0}{upmath}{40}
      \NewMathSymbol{\leqslant}{3}{AMSa}{36}
      \NewMathSymbol{\geqslant}{3}{AMSa}{3E}

      \let\geq=\geqslant \let\ge=\geqslant
    \fi
  \fi
\fi 

\ifnfssone
  \newmathalphabet{\mathit}
  \addtoversion{normal}{\mathit}{cmr}{m}{it}
  \addtoversion{bold}{\mathit}{cmr}{bx}{it}
  \newmathalphabet{\mathbfit} 
  \addtoversion{normal}{\mathbfit}{cmr}{bx}{it}
  \addtoversion{bold}{\mathbfit}{cmr}{bx}{it}
  \newmathalphabet{\mathbfss} 
  \addtoversion{normal}{\mathbfss}{cmss}{bx}{n}
  \addtoversion{bold}{\mathbfss}{cmss}{bx}{n}
  \ifAMStwofonts
    \ifCUPmtlplainloaded \else
      \UseAMStwoboldmath
      \makeatletter
      \new@mathgroup\upmath@group
      \define@mathgroup\mv@normal\upmath@group{eur}{m}{n}
      \define@mathgroup\mv@bold\upmath@group{eur}{b}{n}
      \edef\UPM{\hexnumber\upmath@group}
      \new@mathgroup\amsa@group
      \define@mathgroup\mv@normal\amsa@group{msa}{m}{n}
      \define@mathgroup\mv@bold\amsa@group{msa}{m}{n}
      \edef\AMSa{\hexnumber\amsa@group}
      \makeatother
      \mathchardef\upi="0\UPM19
      \mathchardef\umu="0\UPM16
      \mathchardef\upartial="0\UPM40
      \mathchardef\leqslant="3\AMSa36
      \mathchardef\geqslant="3\AMSa3E

      \let\geq=\geqslant \let\ge=\geqslant
    \fi
  \fi
\fi 

\ifnfsstwo
  \DeclareMathAlphabet{\mathbfit}{OT1}{cmr}{bx}{it}
  \SetMathAlphabet\mathbfit{bold}{OT1}{cmr}{bx}{it}
  \DeclareMathAlphabet{\mathbfss}{OT1}{cmss}{bx}{n}
  \SetMathAlphabet\mathbfss{bold}{OT1}{cmss}{bx}{n}
  \ifAMStwofonts
    \ifCUPmtlplainloaded \else
      \DeclareSymbolFont{UPM}{U}{eur}{m}{n}
      \SetSymbolFont{UPM}{bold}{U}{eur}{b}{n}
      \DeclareSymbolFont{AMSa}{U}{msa}{m}{n}
      \DeclareMathSymbol{\upi}{0}{UPM}{"19}
      \DeclareMathSymbol{\umu}{0}{UPM}{"16}
      \DeclareMathSymbol{\upartial}{0}{UPM}{"40}
      \DeclareMathSymbol{\leqslant}{3}{AMSa}{"36}
      \DeclareMathSymbol{\geqslant}{3}{AMSa}{"3E}

      \let\geq=\geqslant \let\ge=\geqslant
    \fi
  \fi
\fi 

\ifCUPmtlplainloaded \else
  \ifAMStwofonts \else 
    \def\upi{\pi}
    \def\umu{\mu}
    \def\upartial{\partial}
  \fi
\fi

\title[X-ray and soft $\gamma$-ray spectra of Broad-Line Radio Galaxies]
{X-ray and soft $\bmath{\gamma}$-ray spectra of Broad-Line Radio Galaxies}
\author[P. R. Wo\'zniak et al.]
{\parbox[]{6.5in}{Przemys\l aw R. Wo\'zniak$^{1,2}$,
Andrzej A. Zdziarski$^{2}$, David Smith$^3$,\\
Greg M. Madejski$^{4,5}$ and W. Neil Johnson$^6$} \\
 $^1$Dept.\ of Astrophysical Sciences, Peyton Hall, Princeton University,
Princeton, NJ 08544, USA \\
 $^2$N. Copernicus Astronomical Center, Bartycka 18, 00-716 Warsaw, Poland \\
$^3$Dept. of Physics, University of Leicester, University Road, Leicester
LE1 7RH, UK \\
$^4$Lab for High Energy Astrophysics, NASA/GSFC, Greenbelt, MD 20771, USA \\
$^5$Astronomy Dept., University of Maryland, College Park, MD 20742, USA \\
$^6$E. O. Hulburt Center for Space Research, Naval Research Lab, Washington,
DC 20375, USA}
      
\date{Accepted 1998 May 1. Received 1997 April 5}

\pagerange{\pageref{firstpage}--\pageref{lastpage}}
\pubyear{1998}

\begin{document}

\maketitle

\label{firstpage}

\begin{abstract}

We study X-ray and soft \g-ray spectral properties of nearby Broad-Line Radio
Galaxies (BLRGs) using data from \ginga, \asca, OSSE and \exosat. The X-ray
spectra are well fitted by an intrinsic power-law continuum with an energy
index of $\alpha \sim 0.7$, moderately absorbed by a cold medium. In addition,
the \ginga\/ spectra show fluorescent Fe K$\alpha$ lines with an average
equivalent width of $\sim 100$ eV, and, in some cases, Compton reflection
humps. However, the latter components are significantly weaker than both those
seen in radio-quiet Seyferts and those expected if the Fe K$\alpha$ lines were
due to reflection. We find that this weakness of reflection cannot be explained
by dilution by another continuum component, e.g., from a jet. Some \asca\/ and
\exosat\/ spectra show soft X-ray excesses below $\sim 3$ keV. When that
component is taken into account, the Fe K$\alpha$ lines in the \asca\/ data are
found to be unresolved in most cases, and to have equivalent widths $\la 200$
eV, consistent with the \ginga\/ data.

Multiple observations of 3C 390.3 and 3C 382 show the Fe K$\alpha$ line
approximately constant in flux but accompanied by strong continuum variations.
This indicates the bulk of the line is formed by matter at a distance much
larger than an accretion-disk scale, consistent with the \asca\/ line width
measurements. The column density of the matter required to account for the
observed line fluxes is $\nh\ga 10^{23}$ cm$^{-2}$. Such a medium is in the
line-of-sight in 3C 445 but it has to be out of it in other objects, in which
the observed $\nh$ are substantially lower. Thus, a cold medium with that $\nh$
and covering a large solid angle is common in BLRGs but in most object it is
out of the line-of-sight, consistent with the unified AGN model.

The spectra of BLRGs break and become softer above $\sim 100$ keV, as shown by
a simultaneous \asca/OSSE observation of 3C 120 and by the OSSE spectra being
on average much softer than the X-ray spectra. Finally, we find the X-ray and
$\gamma$-ray spectral properties of Cen A, a bright narrow-line radio galaxy --
$\alpha \simeq 0.8$, no or weak Compton reflection, $\nh \ga 10^{23}$ cm$^{-2}$
(which is consistent with the Fe K$\alpha$ line flux), and a high-energy break
at $\sim 100$ keV -- consistent with it being intrinsically very similar to
BLRGs studied here, again in agreement with the unified model.

\end{abstract}


\section{INTRODUCTION}
\label{s:intro}

The 2--20 keV X-ray spectra of Seyfert 1s have been studied using the \ginga\/
data by Nandra \& Pounds (1994, hereafter NP94). Those spectra have been found
not to be simple power laws, but, instead, to show the presence of a
characteristic spectral upturn above $\sim 10$ keV (see also Pounds et al.\
1990), satisfactorily explained by Compton reflection from cold matter,
presumably an accretion disk (Lightman \& White 1988). However, NP94 have not
distinguished in their study radio-quiet Seyferts from radio-loud ones, i.e,
nearby broad-line radio galaxies (hereafter abbreviated as BLRGs). On the other
hand, Zdziarski et al.\ (1995, hereafter Z95) have found that two BLRGs
observed by both \ginga\/ and {\it CGRO}/Oriented Scintillation Spectroscopy
Experiment (OSSE), 3C 111 and 3C 390.3, have the average spectrum with weak or
no Compton reflection. That average spectrum is similar to the intrinsic
spectrum of the nearby narrow-line radio galaxy Cen A, which also does not show
Compton reflection (Miyazaki et al.\ 1996; Warwick, Griffiths \& Smith 1998).

\begin{table*}
\centering
\label{sample}
\caption{The sample of BLRGs. The coordinates, redshifts, and
Galactic column densities, $N_{\rm H,G}$ (in units of $10^{21}$ cm$^{-2}$)
are from Malaguti, Bassani \& Caroli (1994). The NED database was used for
the 5-GHz fluxes, $F_{\rm 5\,GHz}$. The extinction-corrected $B$ magnitudes
were taken from Smith \& Heckman (1989) except for 3C 111, where we used the
$V$ magnitude from Brinkmann et al.\ (1995) and adopted $B-V = 1.7$ from the
AGN catalog of V\'eron-Cetty \& V\'eron (1993).}
 \begin{tabular}{lccccc}
\hline
Object & R.A. & Dec. & $z$ & $N_{\rm H,G}$ & $F_{\rm 5\,GHz}/F_{\rm B}$\\

3C 111     & 04$^{\rm h}$15$^{\rm m}$00$^{\rm s}.4$
& +37$^{\circ}$54$^{\prime}$16$^{\prime \prime}$
&   0.049 &   3.26 &$ 12400$ \\

 3C 120     & 04$^{\rm h}$30$^{\rm m}$31$^{\rm s}.6$
& +05$\degr$15$^\prime$00$^{\prime\prime}$
&   0.033 &   1.23 & 500 \\

 3C 382     & 18$^{\rm h}$33$^{\rm m}$12$^{\rm s}.0$
& +32$\degr$39$^\prime$18$^{\prime\prime}$
&   0.058 &   0.79 & 400  \\

 3C 390.3     & 18$^{\rm h}$45$^{\rm m}$37$^{\rm s}.7$
& +79$\degr$43$^\prime$06$^{\prime\prime}$
&   0.056 & 0.41 & 4300 \\

 3C 445     & 22$^{\rm h}$21$^{\rm m}$14$^{\rm s}.7$
& $-$02$\degr$21$^\prime$25$^{\prime\prime}$
&   0.056 & 0.50 & 1100 \\
\hline
\end{tabular}
\end{table*}

Here we study X-ray and soft \g-ray (hereafter abbreviated as X\g) spectra of
BLRGs observed by \ginga, \asca, OSSE and \exosat. These radio-loud sources can
be classified as Seyfert 1s using optical and UV lines and are listed as such
by NP94. The morphology of most of the objects is consistent with elliptical
(see Section \ref{s:indi}). Our sample contains 5 BLRGs, of which 4 were
observed by \ginga, 5 by \asca, 3  by OSSE, and 5 by \exosat. Table 1 contains
basic information on these objects. We selected the sources having radio to
optical flux ratio $F$(5 GHz)/$F$($B$ optical) $\gg 10$ (Kellermann et al.\
1989).


In Section \ref{s:data} we describe the data used, and in Section
\ref{s:models} we present models used for the subsequent spectral fits.  The
results on individual objects are presented in Section \ref{s:indi}. We
specifically address the issues of the presence of the Compton reflection
component and the strength and width of the fluorescence Fe K$\alpha$ line. In
Section \ref{s:average}, we discuss the average sample properties, the presence
of breaks in soft \g-rays, and differences in X-ray spectral properties between
BLRGs and radio-quiet Seyfert 1s. In Section \ref{s:cena}, we consider the X\g\
spectra of the narrow-line radio galaxy Cen A, and compare them to those of our
sample of BLRGs. In Section \ref{s:dis}, we interpret our results, and
summarize the main conclusions in Section \ref{s:con}.

\section{THE DATA}
\label{s:data}

\begin{table*}
\centering
\label{ginga_log}
\caption{The observation log for \ginga. The count rate is for the top layer
of the LAC in the 1.7--10 keV energy range. The uncertainties here and in
Table 4 below correspond to 1-$\sigma$ confidence intervals.
 }
\begin{tabular}{lcccccc}
\hline
Object & Start date & Start time & End date & End time & Exposure [s] &Count
rate [s$^{-1}$]\\

  3C 111  &  1989-Feb-04 &   13:38:00 &  1989-Feb-05 &   11:00:00 &   18944 &
$12.42 \pm 0.06$ \\

  3C 382  &  1989-Jul-20 &   08:20:00 &  1989-Jul-21 &   00:35:00 &   15488 &
$5.51 \pm 0.08$\\

          &  1989-Jul-21 &   05:40:00 &  1989-Jul-21 &   22:50:00 &   17024 &
$5.13 \pm 0.09$\\

          &  1989-Jul-22 &   06:35:00 &  1989-Jul-24 &   00:00:00 &   30208 &
$5.16 \pm 0.06$\\

3C 390.3  &  1988-Nov-12 &   02:00:00 &  1988-Nov-14 &   03:00:00 &   46720 &
$17.40 \pm 0.06$ \\

          &  1991-Feb-14 &   09:19:00 &  1991-Feb-15 &   02:00:00 &   18176 &
$9.61 \pm 0.06$ \\

  3C 445  &  1988-Nov-02 &   06:40:00 &  1988-Nov-03 &   07:00:00 &   27648 &
$6.92 \pm 0.05$ \\
\hline
\end{tabular}
\end{table*}

\begin{table*} \centering \label{asca_log} \caption{The observation log for
\asca. The count rates for SIS and GIS are given in the 0.55--10 keV energy
band.  The statistical errors of the count rates are $<0.01$ s$^{-1}$, except
for 3C 445, where the error is $<0.002$ s$^{- 1}$.}

\begin{tabular}{lcccccccccc}
\hline
Object & Date & Start time & \multicolumn{4}{c}{Exposure [s]} &
\multicolumn{4}{c}{Count rate [s$^{-1}$]} \\
&&&SIS0 &SIS1 &GIS2 & GIS3 & SIS0 &SIS1 &GIS2 &GIS3 \\
 3C 111 & 1996-Feb-13 &12:35:19 &10200 &32100 &35400 &35400 &0.62 &0.48 &0.50
&0.62 \\
 3C 120 & 1994-Feb-17 &15:42:24 &43500 &43100 &42100 &42100 &1.67 &1.32 &0.98
&1.23 \\
 3C 382 & 1994-Apr-18 &09:24:22 &18300 &18000 &35100 &32700 &1.50 &1.00 &
1.10 &0.84 \\
 3C 390.3 & 1993-Nov-16 &22:40:14 &28800 &31000 &42400 &42400 &0.49 &0.40 &
0.30 &0.39 \\
 & 1995-Jan-15 &09:20:11 &8500 &8600 &13800 &13800 &0.54 &0.46 &
0.36 &0.44 \\
 & 1995-May-06 &00:02:12 &10800 &10800 &13000 &13000 &0.96 &0.81 &
0.62 &0.72 \\
 3C 445 & 1995-Jun-01 &11:43:49 &33600 &34200 &35300 &35300 &0.057 &0.048
&0.046 &0.063 \\
\hline
\end{tabular}
\end{table*}

\begin{table*}
\centering
\label{t:osse_log}

\caption{The observation log for OSSE. VP stands for the Viewing Period of {\it
CGRO}. The net exposure time and the count rates are normalized to a single
OSSE detector. The count rates and the photon fluxes (in units of
$10^{-4}$\,s$^{-1}$ cm$^{-2}$) are given for the 50--150 keV band, and the
energy spectral index, $\alpha$, for the 50--500 keV band. }

\begin{tabular}{lccccccc} \hline Object & VP &Start date &End date & Exposure
[s] & Count rate [$10^{-2}$ s$^{-1}$]& Photon flux  & $\alpha$ \\

3C 111 &4& 1991-Jun-28 & 1991-Jul-12 & 440,233 & $15.04\pm 2.52$ & $3.20\pm
0.60$ & $1.04 \pm
0.34$  \\

&29& 1992-May-14 & 1992-Jun-04 & 184,973 & $5.64\pm 3.67$ & $1.86\pm 1.12$ &
$-0.21\pm 1.06$
\\

&Sum& 1991-Jun-28 & 1992-Jun-04 & 625,206 & $12.26\pm 2.08$ & $2.81\pm 0.54$ &
$1.00\pm 0.35$
\\

3C 120  & 29 & 1992-May-14 & 1992-Jun-04  & 162,711 & $9,19\pm 3.91$ & $2.34\pm
1.21$ & \\

& 30 & 1992-Jun-04 & 1992-Jun-11  & 154,046 & $<9.22$ & $<2.21$ &\smash{\raise
0.25cm\hbox{$0.63 \pm 0.79$}} \\

& 33 & 1992-Jul-02 &  1992-Jul-16 & 298,222 & $19.75\pm 3.30$ & $4.60\pm 0.78$
& $2.02 \pm
0.50$ \\

& 220& 1993-May-08 &  1993-May-13 & 94,259 & $17.68\pm 5.00$ & $3.91\pm 1.16$ &
$1.25 \pm 0.81$
\\

& 224& 1993-Jun-04 &  1993-Jun-14 & 459,622 & $15.12\pm 2.23$ & $3.41\pm 0.52$
& $1.20 \pm
0.34$   \\

& 317 & 1994-Feb-17 & 1994-Mar-01 & 208,365 & $8.16\pm 4.29$ & $1.63\pm 1.01$ &
$0.55 \pm
0.83$   \\

& 320 & 1994-Mar-08 & 1994-Mar-15 & 201,880 & $8.11\pm 4.53$ & $2.05\pm 1.07$ &
$4.10 \pm
2.28$  \\

& 617.1 & 1997-Mar-18 & 1997-Apr-01 & 669,576 & $9.01\pm 2.72$ & $ 1.96\pm
0.65$ & \\

& 617.7  & 1997-Apr-07 & 1997-Apr-09 & 109,300 & $13.77\pm 7.38$ & $3.28\pm
1.76$ & \smash{\raise 0.25cm\hbox{$0.68\pm 0.58$}} \\

& Sum & 1992-May-14 & 1997-Apr-09 & 2,357,981 & $11.22\pm 1.25$ & $2.54\pm
0.30$ & $1.10 \pm
0.26$ \\

3C 390.3 & 12 & 1991-Oct-18 &  1991-Oct-31 & 375,313 & $18.07\pm 2.92$ &
$4.04\pm 0.69$ &$1.38
\pm 0.43$ \\

& 29 & 1992-May-15 &  1992-Jun-04 & 390,664 & $11.55\pm 2.52$ & $2.58\pm 0.60$
& $1.19 \pm
0.45$ \\

& 209 & 1993-Feb-10 &  1993-Feb-22 & 330,724 & $5.69\pm 3.13$ & $1.26\pm 0.74$
& $1.19\pm 1.08$  \\

&Sum & 1991-Oct-18 &  1993-Feb-22 & 1,096,700 & $12.01\pm 1.64$ & $2.54\pm
0.40$ & $1.36 \pm
0.32$   \\

Sum&& 1991-Jun-28 &  1997-Apr-09 & 4,079,887 & $11.34\pm 0.91$ & $2.57\pm 0.22$
& $1.15 \pm 0.16$  \\
 \hline
\end{tabular}
\end{table*}

For our study, we have selected BLRGs detected by either \ginga, \asca, or
OSSE. In addition, we consider observations of those objects by \exosat.

A total of 5 radio-loud Seyfert 1s were observed by \ginga\/ LAC (the numbers
of individual observations are given in brackets): 3C 98 [1], 3C 111 [1], 3C
382 [3], 3C 390.3 [2], and 3C 445 [1]. A rigorous data selection policy was
adopted in order to exclude periods of high or unstable background (see, e.g.,
Smith \& Done 1996). 3C 98 has not been detected and thus it is not included in
our analysis. Table \ref{ginga_log} contains the observation log for \ginga.
The observations of 3C 111, 3C 382 and of the 1988 observation of 3C 390.3 are
included in the sample of NP94, and of 3C382, 3C390.3, and 3C 445 in the sample
of Lawson \& Turner (1997).

The background was estimated using one of the two methods given by Hayashida et
al.\ (1989) (see also Williams et al.\ 1992). The local method uses off-source
observations taken within two days of (or at the same orbital phase as) the
on-source observation to estimate the background level at the time of the
observation, while the universal method uses all the off-source observations
taken within a contemporary four month period to model systematic trends in the
particle background levels, and hence to estimate the background level at the
time of the source observation. It is found that the universal method gave poor
results for one observation of 3C 390.3 and all observations of 3C 382. For the
remaining observations we used the results obtained from the universal method
as it minimises the fluctuations in the cosmic X-ray background (Hayashida et
al.\ 1989; Williams et al.\ 1992). For 3C 382 and 3C 445, the weakest of the
sample, we have also added and subtracted a power law with $\alpha=0.8$ with
the 1-keV normalization of $5\times 10^{-4}$ cm$^{-2}$ s$^{-1}$ (representing
background fluctuations) to the data. The effect on the spectral slope below 10
keV was small, $\pm 0.05$, and completely negligible above 10 keV. Thus, we
ignore this effect in fitting the data. Finally, due to the low Galactic
latitude of 3C 111, the \ginga\/ observation is possibly contaminated by the
soft diffuse X-ray emission from our galaxy. To overcome this, we subtracted
the best estimate of the local soft X-ray emission (derived from a nearby
off-source observation) from the source data.

Data were then extracted from both the top-layer and mid-layer of the LAC, as
the mid-layer has more effective area above $\sim 10$ keV (although we ignore
the mid-layer data below 10 keV as these are subject to greater uncertainties
in background estimation). Contamination from fluorescence by silver atoms in
the LAC collimator complicates the spectral analysis around $\sim 22$ keV; for
this reason, our analysis is restricted to the $\sim 1.7$--18 keV energy
range. Prior to spectral fitting, the data were corrected for the collimator
response and a 0.5 per cent systematic error was added to each {\sc pha}
channel, to account for uncertainties in the detector response.

\asca\/ satellite consists of four co-aligned X--ray telescopes, and it covers
a bandpass roughly from 0.4 to 10 keV (see Tanaka, Inoue \& Holt 1994). The
focal planes of two of the four telescopes have Gas Scintillation Imagers
(GISs), while the remaining two have X--ray sensitive CCD cameras (Solid-state
Imaging Spectrometers, or SISs), each consisting of four CCD detectors
(chips).  The observations of 3C 111 [1], 3C 120 [1], 3C 390.3 [3] and 3C
445 [1] were performed in a standard spectroscopy configuration with the
source focussed on chip 1 in the SIS0 and on chip 3 in the SIS1.  The
observation of 3C 382 was performed with the source focussed on chip 2 in
the SIS0 and on chip 0 in the SIS1 (a somewhat non-standard configuration).
The log of the \asca\/ observations is given in Table 3.

The data were extracted from the HEASARC archives, and reduced using standard
\asca\/ data reduction procedures. We used the \asca\/ rev.\ 2 processing,
extracting manually the event files, but in the process of doing so, we used
the standard screening criteria.  In all cases, the SIS data presented here
were analyzed in the highest spectral resolution ``faint'' mode, to assure that
there would be no additional gain offset from the potential contamination by
the scattered sunlight.  For 3 objects, this results in a slightly lower
effective observation time and thus slightly lower statistics than in the
``bright'' mode. However, there was no discernible difference between the
results of the ``faint'' and ``bright'' mode spectral fits, as discussed in
Section 4 below.  On the other hand, the use of the ``faint'' mode for 3C 111
and 3C 382 resulted in only about 40 per cent of SIS0 data and 50 per cent of
all SIS data, respectively, as compared to the ``bright'' mode. Therefore, in
addition to the ``faint" mode, we also performed the spectral fits to the data
analyzed in the ``bright'' mode for those 2 objects, see Sections 4.2 and 4.4.

The resulting spectra were then binned to assure that there are $\geq 20$
counts per bin.  In addition, for the observation of 3C 120, due to an on-board
electronics malfunction in the GIS3 -- where the two least significant bits of
the analog-to-digital converter in the Pulse Height Discriminator circuit were
stuck in a fixed pattern -- all data for GIS3 needed to be binned with at least
8 channels to a bin.  In all cases, the source photons were extracted using
circular regions, with radii of $3^\prime$ for the SISs, and $6^\prime$ for the
GISs, while the background, to be subtracted prior to the spectral fits from
the source counts to get the net spectra, was extracted from suitable
source-free regions of a size comparable or greater than the source regions.

For the GIS data, in our spectral fitting, we used the standard GIS response
matrices (v4.0).  For the SIS data, we used the SIS response matrix generator
({\sc sisrmg}), as appropriate to every observation and detector, v1.1, April
1997.  For both SIS and GIS data, we used the telescope effective areas (via
the use of {\sc ascaarf} v2.72, March 1997).  To account for the residual
errors in the relative cross-calibration of the four telescopes, we fit the
same model to the four data sets with each normalization as a free
parameter. For plotting purpose only, we further rebin, renormalize and add the
data from all 4 detectors using the normalization of SIS0. That normalization
is also used for comparison with results of other instruments. We note that the
relative normalization differences in our data with respect to SIS0 are $<13$
per cent except for the 3C382 observation (which used the non-standard
observation configuration as described above) where the differences are $<26$
per cent.

Most of the OSSE data used here have been briefly reported in Johnson et al.\
(1997). The analysis of 3C 111 [2], 3C 120 [9] and 3C 390.3 [3] takes into
account a systematic error correction to the spectra computed from the
uncertainties in calibration and response of the detectors using both in-orbit
and pre-launch data. The systematic errors correspond to $\sim 3$ per cent
uncertainty in effective area at 50 keV, decreasing to $\sim 0.3$ per cent at
150 keV and above. We use the OSSE response matrix as revised in 1995, which
results in the 50--60 keV fluxes about 20 per cent higher than in the original
response (used, e.g., in Z95). Observation log for OSSE is given in Table 4.

We supplement the above data set by selected data from \exosat. We use
\exosat\/ ME spectra from the HEASARC archive with the quality flag 3 or
higher, which indicates observations with reliable background subtraction (note
that this criterion removes 3C 111 from the sample). The objects are 3C 120
[14], 3C 382 [24], 3C 390.3 [4], and 3C 445 [1] (some of those observations are
reported in Turner \& Pounds 1989). The usable \exosat\/ energy range is from
1.2 keV to 8 keV in most cases as the spectra above 8 keV suffer from
relatively inaccurate background subtraction. The individual spectra include a
1 per cent systematic error. Since the individual spectra are of rather low
statistical significance, we use co-added spectra of each of the first 3
objects with the weights corresponding to the length of time of each
observation. Both the counts and the response matrices for each observation
were added using the {\sc addspec} function of the {\sc ftools} data processing
package. Time intervals covered by the observations are 1983 March--1986
February, 1983 September--1985 September, 1985 February--1986 March, and 1984
May for 3C 120, 3C 382, 3C 390.3 and 3C 445, respectively. We note that the
\fek\ line energies in the \exosat\/ data appear at energies significantly
lower than those obtained with \ginga\/ and \asca, which appears to be due to a
calibration problem.

\section{SPECTRAL MODELS}
\label{s:models}

We use {\sc xspec} (Arnaud 1996) v9 and 10 for spectral fitting. The parameter
uncertainties in Tables below correspond to 90 per cent confidence for a single
parameter, i.e., $\Delta \chi^2=+2.7$. On the other hand, the plotted error
bars are 1-$\sigma$, the upper limits, 2-$\sigma$, and the spectral data are
rebinned for clarity of display.

Fitted intrinsic spectra are absorbed by intervening matter with the column
density that consists of both the Galactic component, $N_{\rm H, G}$ (see Table
\ref{sample}), and the column density intrinsic to the source, $\nh$ (note that
$\nh$ does not include $N_{\rm H,G}$). Both the intrinsic spectra and
absorption by $\nh$ are evaluated at the source redshift, $z$. The absorption
is in neutral matter with the abundances of Anders \& Ebihara (1982) and the
opacities of Morrison \& McCammon (1983), as implemented in the {\tt zwabs}
model of {\sc xspec}.

We model the underlying continuum as a power law with an energy spectral index,
$\alpha$, multiplied by an exponential with an e-folding energy, $\ec$. When
fitting the X-ray data only, we assume $\ec=400$ keV, consistent with results
of Z95. This, however, has a negligible effect on the resulting parameters. On
the other hand, $\ec$ is a free parameter in fits to combined X\g\ data.

We also allow for the presence of a Compton-reflection spectral component
(Lightman \& White 1988). Compton reflection arises
when the underlying component irradiates cold matter in the vicinity of the
nucleus, e.g., an accretion disk or a torus. We use Green's functions for
angle-dependent reflection of isotropic incident radiation of Magdziarz \&
Zdziarski (1995), and assume the viewing angle of $30\degr$ (corresponding to
an orientation close to face-on, as expected in type-1 AGNs, Antonucci 1993,
see also Eracleous, Halpern \& Livio 1996 for 3C 390.3). The relative amount of
reflection is measured by the solid angle subtended by the reflector, $\Omega$
($\Omega=2\pi$ corresponds to reflection from an infinite slab). The opacities
of reflector are the same as for the absorber. However, we consider some models
with the Fe abundance with respect to that of Anders \& Ebihara (1982), $\af$,
being $>1$. In some cases, we also allow the reflecting medium to be ionized,
in which case we use the model of Done et al.\ (1992) modified as in Gondek et
al.\ (1996). We assume the reflector temperature of $10^5$ K (Krolik \& Kallman
1984). The ionization parameter is defined as $\xi\equiv L_{\rm ion}/n r^2$,
where $L_{\rm ion}$ is the luminosity in an incident power-law continuum in the
5 eV--20 keV range, $n$ is the density, and $r$ the distance between the source
of radiation and the reflector. The neutral and ionized reflection is
implemented in the {\sc xspec} models {\tt pexrav} and {\tt pexriv},
respectively.

Both reflection and absorption of the continuum are accompanied by emission of
fluorescent lines, the most prominent of those is the Fe K$\alpha$ line (at 6.4
keV in the case of neutral Fe), which we model as a Gaussian at an energy,
$\efe$, with a width, $\sfe$, and a photon flux, $\ife$. The line equivalent
width is denoted by $\wfe$. The line flux is allowed to be a free parameter,
independent of both the amount of reflection and absorption (unless stated
otherwise in the text). In fitting the data from \ginga\/ and \exosat, we
assume the fixed $\sfe=0.1$ keV. On the other hand, $\sfe$ is a free parameter
in fits to the \asca\/ data, which have energy resolution much better than
those of \ginga\/ and \exosat. In fits to \asca\/ data, we include, in addition
to the K$\alpha$ line, the Fe K$\beta$ and Ni K$\alpha$ lines, a sum of which
we model as a single line with the energy, width, and normalization equal 1.1,
1.1, and 0.2, respectively, of the corresponding values for Fe K$\alpha$ (e.g.,
\.Zycki \& Czerny 1994). For simplicity, we denote below both lines as \fek.
The line parameters given below for \asca\/ data correspond to the Fe K$\alpha$
component only. (Note that for clarity of plotting \asca\/ data, we use
$\sfe=0.1$ keV whenever fitting yields a lower value.) Except for 3C 445 and
Cen A, we assume the lines are absorbed in the same way as the continuum.

\section{RESULTS OF SPECTRAL FITS}
\label{s:indi}

\subsection{3C 445}
\label{ss:3c445}

\begin{table*}
\centering \label{t:3c445}

\caption{Results of fits to the 3C 445 data from \ginga, \asca\/ and \exosat.
The \ginga\/ and \exosat\/ data contain a cluster contribution at low energies,
which is modeled by a 3-keV bremsstrahlung  photon spectrum, $A_2\exp(-E/kT)$
times the Gaunt factor. $A_1$ gives the 1-keV normalization of the power-law
component.  The \asca\/ continuum is modeled as the sum of 3 power-law
components with the same $\alpha$ but different normalizations, $A_i$, and
columns, $N_{{\rm H},i}$.  Hereafter, the normalizations are given in units of
$10^{- 3}$ s$^{-1}$ cm$^{-2}$, $\nh$ is given in $10^{21}$ cm$^{-2}$, $\efe$
and $\sfe$ in keV, $\ife$ in $10^{-5}$ s$^{-1}$ cm$^{-2}$, $\wfe$ in eV, and
`f' denotes a fixed parameter.   }

\begin{tabular}{ccccccccc}
\hline
$A_i$ & $N_{{\rm H},i}$ & $\alpha$ & $\Omega/2\pi$ & $\efe$ & $\sfe$ & $\ife$ &
$\wfe$ & $\cd(\cnu)$ \\
 \multicolumn{9}{c}{\ginga\/ 1988 Nov} \\
$3.2_{-0.6}^{+1.7}$, $4.3_{-0.2}^{+0.2}$ & $150^{+30}_{-30}$, 0f &
$0.45^{+0.14}_{-0.10}$ & $0^{+0.22}$ & $6.37^{+0.22}_{-0.19}$ &0.1f
& $4.7^{+1.7}_{-2.1}$ & $150$ & 35.3/36(0.98) \\
 \multicolumn{9}{c}{\asca\/ 1995 Jun} \\
$1.9^{+2.6}_{-1.2},0.87^{+0.54}_{-0.39},0.15^{+0.02}_{-0.02}$ &
$580^{+370}_{-
230},78^{+21}_{-20},0^{+0.60}$ & $0.39^{+0.28}_{-0.29}$ & 0f &
$6.44^{+0.07}_{-0.07}$ & $0^{+0.18}$ & $1.8^{+0.6}_{-0.7}$ & 140 &
429/421(1.02) \\
 \multicolumn{9}{c}{\exosat\/ 1984 May} \\
$2.5^{+9.0}_{-1.4},\,0.62^{+0.87}_{-0.62}$ & $80^{+100}_{-50}$, 0f &
0.$33^{+0.77}_{-0.40}$ & 0f & -- & -- & 0f & -- & 35.8/43(0.83) \\
 \hline \end{tabular}
\end{table*}

3C 445 is a powerful FR II BLRG with a typical lobe-dominated radio morphology
and an elliptical appearance in the optical band (Smith \& Heckman 1989).  The
source lies $\sim 0.5\degr$ from the cluster A 2440, which thus contaminates
the spectrum of 3C 445 (Pounds 1990) from \ginga\/ (which has the field of view
of $1\degr \times 2\degr$).  The cluster emission has been modelled by Pounds
(1990) as thermal bremsstrahlung at $kT=3$ keV.  We note that component would
also include any soft X-ray excess of the AGN itself.  We fit the updated
\ginga\/ data (see Section \ref{s:data}) by the sum of of that component, an
absorbed power law with Compton reflection and an \fek\ line, and obtain the
parameters given in the Table 5.  They are consistent with those of Pounds
(1990) within the statistical uncertainties, although Pounds (1990) obtains a
somewhat softer spectral index, $\alpha=0.68^{+0.20}_{-0.18}$, which is
apparently due to differences in data processing.  Similarly to Pounds (1990),
we find the X-ray source is strongly obscured.  The \ginga\/ spectrum is shown
in Figure 1a.

\begin{figure}
\label{fig:3c445}
\centering
\epsfxsize=8.4cm \epsfbox{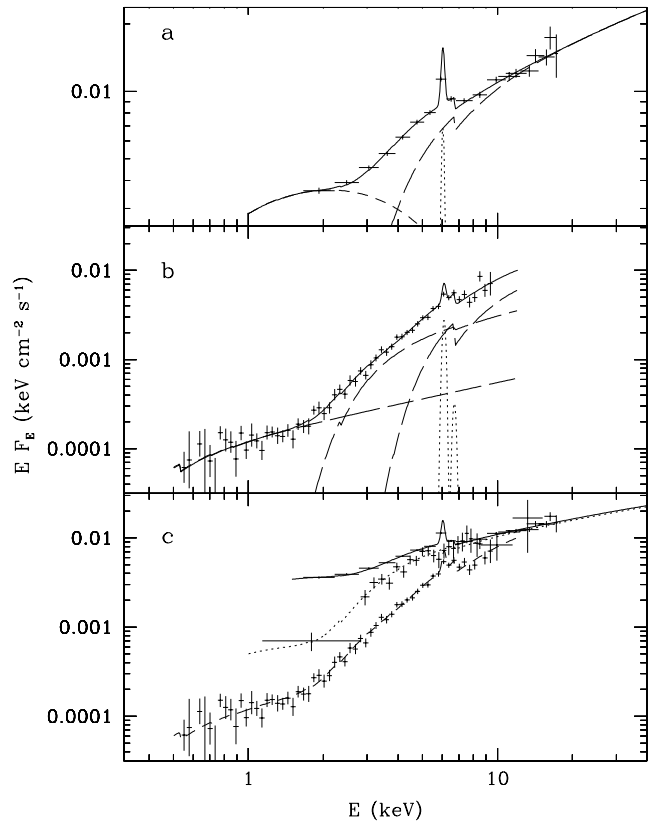}

\caption{The X-ray spectra (crosses) of 3C 445. {\it (a)} The data from \ginga,
with emission from the nearby cluster A 2440 modeled as bremsstrahlung (short
dashes), an absorbed power-law emission of 3C 445 (long dashes), and an \fek\
line emission (dots) from both the AGN and the cluster.  The solid curve is the
sum.  {\it (b)} The data from \asca\/ (which contain no contribution from the
cluster).  The spectrum is modeled as the sum (solid curve) of 3 power-law
components (long dashes) with the same $\alpha$ but different $\nh$, and an
\fek\ line (dots).  {\it (c)} Comparison of the spectra from \ginga, \exosat,
and \asca\/ (from top to bottom). } \end{figure}

We see in Table 5 that the \ginga\/ data do not require the presence of a
Compton reflection component, and $\Omega/2\pi \la 0.2$. (The top-layer
\ginga\/ data alone are consistent with this result, but yield a weaker
constraint of $\Omega/2\pi= 0.20_{-0.20}^{+0.64}$.) On the other hand, there is
a strong \fek\ line in the spectrum (modeled here as absorbed by $N_{\rm H,G}$
only).  The line originates in both the cluster and the AGN.  The cluster emits
a thermal emission line at the rest-frame energy of 6.7 keV with $\wfe \simeq
500$ eV with respect to the bremsstrahlung spectrum (Pounds 1990 and references
therein).  At $z = 0.094$ of A 2440 (Struble \& Rood 1987), the observed
centroid energy of the cluster line is 6.12 keV and cannot be resolved by
\ginga\/ from the fluorescent line from 3C 445 redshifted from 6.4 keV to 6.05
keV.  Thus, we model both line components as a single Gaussian.  The line flux,
$\ife$, in Table 5 corresponds to both components, whereas the stated $\wfe$
corresponds to the AGN component only with respect to the absorbed power-law
continuum.

The \asca\/ data, on the other hand, are basically free from contamination from
the cluster.  In our modeling of the continuum, we use a model with 3 power-law
components each absorbed by a different column density, which was shown to be
the simplest model accounting for the shape of the spectrum from \asca\/ in an
early analysis of the data by Yamashita \& Inoue (1996).  The fit parameters
are given in Table 5, and the spectrum is shown in Figure 1b.  We see that the
spectrum of 3C 445 contains a substantial soft X-ray excess component, modeled
here as the third power-law component with $\nh$ consistent with null (Table
5).  That component can be interpreted as due to scattering of about 5 per cent
of the intrinsic emission by a hot plasma in the funnel of an obscuring torus
(Krolik \& Kallman 1987).  Then the remaining two absorbed power-law components
can be interpreted as being due to the product of uniform absorption with
$N_{\rm H,2}$ and partial covering by $(N_{\rm H,1}-N_{\rm H,2})$ with a
covering fraction of $A_1/(A_1+A2)$.

The obtained spectral index and the line parameters are similar to those
derived above from the \ginga\/ data.  In particular, the obtained line
equivalent width is almost the same as $\wfe$ estimated above after
subtracting the cluster component in the \ginga\/ data.  We also find the
\fek\ line is narrow, $\sfe<0.18$ keV.  The $\sfe$-$\efe$ confidence contours
are presented in Figure 2, and the line profile is shown in Figure 3.  Note
that the presence of a soft excess component in the spectrum of 3C 445
shown by \asca\/ implies that a (minor) part of the soft excess (modeled as
bremsstrahlung) in the \ginga\/ and \exosat\/ data actually comes from
the AGN itself.

Our \asca\/ results can be compared with a recent independent analysis of the
data by Sambruna et al.\ (1998). Their best-fit model has the identical form to
ours. However, they obtain a harder power law than that obtained by us, with
$\alpha=0.25^{+0.29}_{-0.27}$, and a broader and stronger \fek\ line, with
$\sfe=0.16^{+0.16}_{-0.13}$ keV (in the source frame), and $\ife=
(5.0^{+4.0}_{-2.5}) \times 10^{-5}$ s$^{-1}$ cm$^{-2}$. These differences can
be explained by their use of an earlier (of 1994) version of the \asca\/
software than that used by us (of 1997). Furthermore, the reduced $\chi^2$ of
their best-fit model is statistically significantly larger than ours (1.05 vs.\
1.02). Thus, we conclude that the \asca\/ data are fully compatible with the
\fek\ line being narrow, as obtained by us.

\begin{figure} \label{fig:contour}
\begin{center}
\leavevmode
\epsfxsize=8.4cm\epsfbox{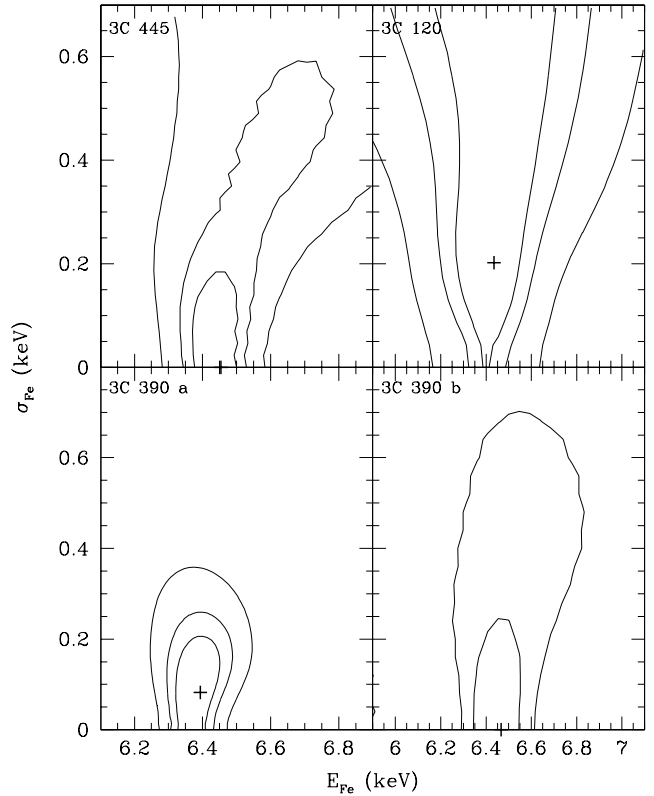}
\end{center}

\caption{The confidence contours (at 68, 90, and 99 per cent) for the energy
and the width of the iron line (in the source frame) for the \asca\/
observations showing statistically significant lines. Models with Compton
reflection are used except for 3C 445, see Tables 5, 7, 9. } \end{figure}

\begin{figure*}
\label{fig:ratio}
\begin{center}
\leavevmode
\epsfxsize=11cm \epsfbox{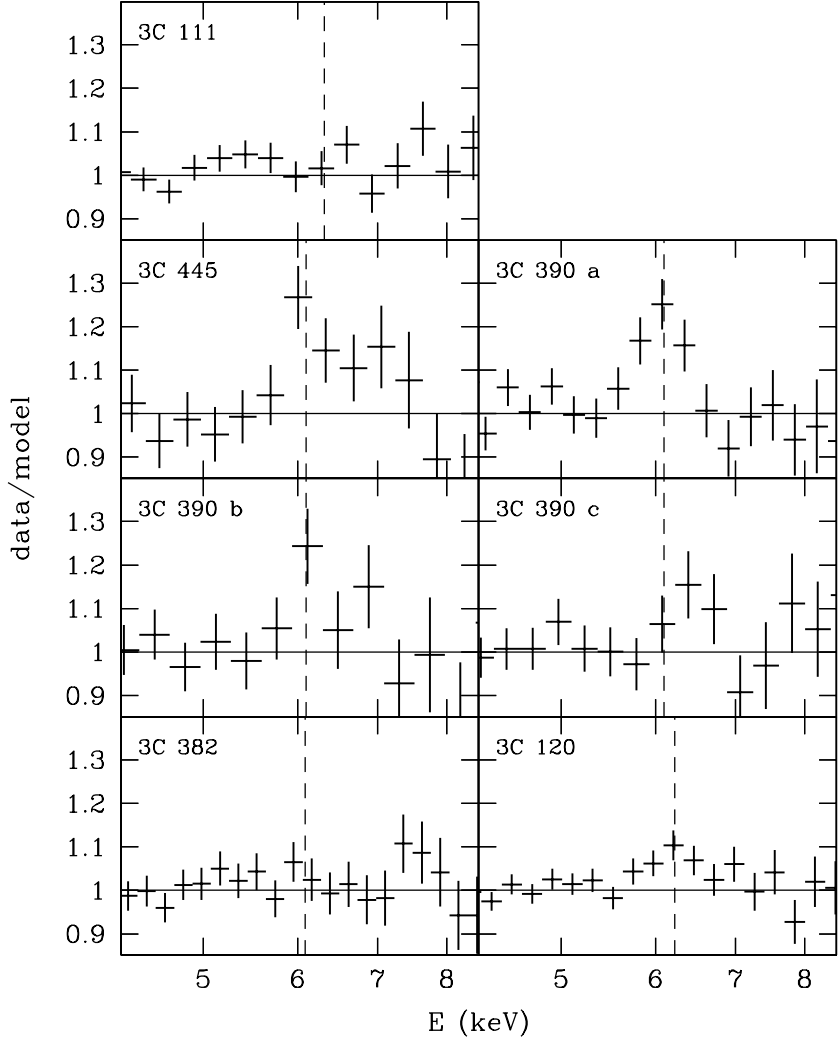}
\end{center}

\caption{The line profiles for all presented \asca\/ observations, modeled
including Compton reflection except for 3C 445, see Tables 5-9. The data have
been fitted including the \fek\ line, which then was removed from the model in
the displayed ratios.  The vertical dashed lines correspond to 6.4 keV at the
redshift of each source.  } \end{figure*}

The $\wfe\sim 150$ eV obtained by us for both \ginga\/ and \asca\/ data is
consistent with emission of a shell of cold matter with $\nh\sim
(2$--$5)\times 10^{23}$ cm$^{-2}$ and the cosmic Fe abundance irradiated
isotropically from the center (Makishima 1986, hereafter M86; Awaki et al.\
1991).  We thus conclude that the most likely origin of the \fek\ emission in
3C 445 is fluorescence in the observed absorber, which, we infer, covers a
large solid angle around the nucleus.  This interpretation is consistent with
the narrowness of the line derived from the \asca\/ data.  The presence of an
optically-thick disk irradiated isotropically by the X-ray source appears to be
ruled out as the solid angle of any reflector is constrained to $\ll 2\pi$.

Figure 1c shows a comparison of 3C 445 spectra from different instruments.
Here, we also show the spectrum from \exosat, fitted in the 1--15 keV range in
the same way as the \ginga\/ data except that the \fek\ line is not included in
the model due to the limited statistical quality of the former data. The
cluster contribution to the \exosat\/ data is about a quarter of that in the
\ginga\/ data, which is fully consistent with the field of view of \exosat\/
($45'\times 45'$) being smaller than that of \ginga. On the other hand, the
spectra of the AGN itself from \ginga\/ and \exosat\/ are consistent with being
the same. However, the \asca\/ spectrum lies significantly below the other two
spectra even at hard X-ray energies, where the cluster contribution is
negligible. This indicates the presence of X-ray variability in 3C 445, caused
by either variability of the power-law component or variable absorption.

\subsection{3C 111}
\label{ss:3c111}

\begin{table*}
\centering
\label{t:3c111}

\caption{Results of fits to the \ginga\/ and \asca\/ spectra of 3C
111.}

\begin{tabular}{lcccccccc}
\hline
$A$ & $\alpha$ & $\nh$ & $\Omega/2\pi$ & $\efe$ & $\sfe$ & $\ife$  &
$\wfe$ & $\cd(\cnu)$ \\
\multicolumn{9}{c}{\ginga\/ 1989 Feb}\\
$10.5_{-0.09}^{+0.14}$ & $0.82^{+0.08}_{-0.06}$ & $18^{+2}_{-2}$ &
$0.12^{+0.31}_{-0.12}$ & $6.72^{+0.36}_{-0.32}$ & 0.1f &
$2.7^{+1.7}_{-1.5}$ & 78 & 36.8/34(1.08) \\
 \multicolumn{9}{c}{$\ife/\Omega =$ const} \\
11.3 & $0.86^{+0.06}_{-0.07}$ & $19^{+2}_{-2}$ &
$0.30^{+0.23}_{-0.21}$ & $6.79^{+0.51}_{-0.46}$ & 0.1f &
$2.0^{+1.5}_{-1.4}$ &   $55$ & 39/35(1.10) \\

\multicolumn{9}{c}{\asca\/ 1996 Feb} \\
$9.6^{+0.4}_{-0.3}$ & $0.73^{+0.03}_{-0.02}$ & $7.3^{+0.3}_{-0.3}$ & 0.12f
& $6.78^{+\infty}_{-6.78}$ & 0.1f & $1.0^{+1.4}_{-1.0}$ & 25 & 1418/1511(0.94)
\\
\hline
\end{tabular}
\end{table*}

3C 111 is a luminous FR II radio galaxy with bright core, a one-sided jet on
milli-arcsecond scale (Linfield \& Perley 1984), a highly asymmetric double
radio structure, and reported superluminal motion ($\ga 10 c$,  Preuss,
Alef \& Kellermann 1988). The optical image shows no structure (Colina \&
P\'erez-Fournon 1990) and thus exact classification is not possible, but the
properties of the host galaxy are consistent with elliptical morphology.

\begin{figure}
\centering
\epsfxsize=8.4cm \epsfbox{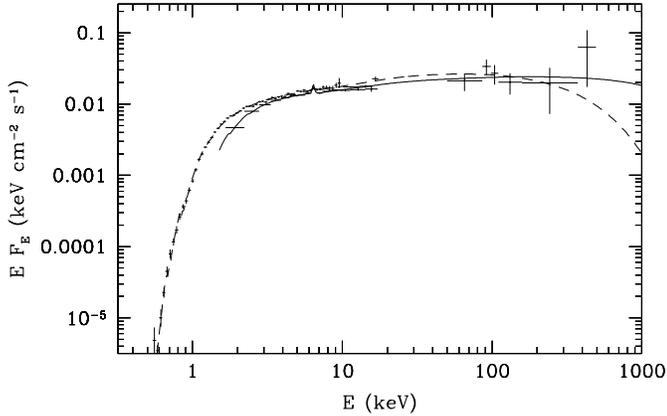}
\label{fig:3c111}

\caption{The spectrum of 3C 111 from the 1989 observation by \ginga\/ (the
bottom X-ray spectrum), the 1996 observation by \asca\/ (the top X-ray
spectrum), and the 1991-92 OSSE spectrum ($>50$ keV).  } \end{figure}

The spectrum of 3C 111 from \ginga\/ was included in the sample of NP94. The
parameters of our fits to the \ginga\/ observation are reported in Table 6 and
the spectrum is shown in Figure 4.  The \fek\ line energy, $\sim 6.7$ keV
(although lower than 7.4 keV found by NP94), suggests emission by strongly
ionized iron. The remaining fit parameters are similar to those of NP94, in
particular there is little evidence for reflection. (The top-layer \ginga\/
data alone are also consistent with the absence or weakness of reflection,
yielding $\Omega/2\pi= 0^{+0.16}$.) On the other hand, the best-fit value of
$\wfe\simeq 80$ eV appears too large for the \fek\ line to be due to
reflection. However, the uncertainties on the fitted parameters are such that
the origin of the \fek\ line from Compton reflection is still possible. In
particular, we have fitted a model in which the line equivalent width was tied
to the reflected component. Theoretical estimates of $\wfe$ with respect to the
Compton-reflected component are relatively uncertain and vary from $\sim 0.7$
keV to $\sim 1.5$ keV for $\Omega=2\pi$, $\al = 0.9$ and $\af = 1$ (\.Zycki \&
Czerny 1994; George \& Fabian 1991; hereafter GF91). The line becomes stronger
with increasing Fe abundance (GF91). Here, we adopt the highest value of 1.5
keV for $\af=1$ and increase it with increasing $\af$ according to the results
of GF91. We find that already for $\af=1$, $\Omega/2\pi \sim 0.3$ at $\Delta
\chi ^2 \simeq +2$ (Table 6) which increase is not significant statistically.
This means that the line emission from 3C 111 can originate from Compton
reflection, but the $\Omega/2\pi \sim 0.3$ implies the reflector geometry is
different from a slab.

We have tested whether ionization of the reflector can relax the constraints on
the solid angle obtained above. We have found that the \ginga\/ data cannot
constrain the ionization parameter at all, but the best-fit $\xi=0$. When we
fix $\xi$ at a large value, e.g., $\xi=5000$ erg s$^{-1}$ cm$^{-1}$, the
allowed solid angle of the reflector {\it decreases\/} with respect to neutral
reflection: $\Omega/2\pi=0^{+0.09}$. This decrease is a consequence of a larger
depth of the K edge from an ionized medium than a neutral one with the data
consistent with no K edge. Thus, the small values of $\Omega/2\pi$ obtained
here are not an artefact of the assumption of the neutrality of the reflector.
On the other hand, the reflected X-ray spectrum would have the same shape as
the incident spectrum in the limit of complete ionization of all elements in
the reflector (White, Lightman \& Zdziarski 1988), which would then allow any
value of $\Omega$, in particular $2\pi$. However, such strong ionization would
lead to {\it no\/} Fe K$\alpha$ line.

In fitting the \asca\/ data, we assumed the best-fit $\Omega/2\pi$ obtained
from the \ginga\/ data, which, due to the weakness of Compton reflection,
affects little the fit. We find only a weak \fek\ line in these data, with
$\wfe\sim 25^{+30}_{-25}$ eV, see Table 6 and Figure 3.  Due to its weakness,
the line parameters are not constrained, and $\sfe$ has been fixed at 0.1 keV
while determining the confidence contours of other parameters.

We have investigated whether we could somehow miss a broad and strong \fek\
line still present in the \asca\/ spectrum. First, we have found that the line
disappears altogether from the model if Compton reflection is assumed to be
strong, $\Omega/2\pi= 1$, which $\Omega$ is expected in the case of line
formation in an inner region of an accretion disk. Second, we have considered
the effect of using  the SIS0 data obtained in the bright mode, which increases
the exposure available for that detector by a factor of about 2.5 (but degrades
the energy resolution, see Section 2). We find that then the \fek\ line in the
combined SIS/GIS data becomes only slightly stronger, with the best-fit
$\wfe\simeq 35$ eV. Again, the assumption of $\Omega/2\pi=1$ results in
disappearing of the line. If the SIS0 and SIS1 data are used without the GIS
data (which have a worse energy resolution), $\sfe\simeq 0.2_{-0.2}^{+1.1}$
keV, and the best-fit $\wfe\simeq 60$--90 eV are obtained for $\Omega/2\pi$
between 1 and 0.12. Thus, any \fek\ line in the data is at most moderate and
its strong broadening is neither required nor implied by the data.

The \asca\/ data show little intrinsic spectral variability with respect to the
\ginga\/ data, see Table 6 and Figure 4.  However, the \asca\/ data show a
marked decrease of the absorbing column with respect to that seen by
\ginga. This decrease correlates with the apparent decrease of the \fek\ line
flux between the \ginga\/ and \asca\/ observations, which suggests the origin
of the line primarily in the absorber. Although the best-fit value of $\wfe$
from \ginga\/ is a factor of $\sim 3$ larger than that expected from a shell of
absorbing, isotropically-irradiated, matter with the best-fit value of $\nh$
(M86), the uncertainties on the line flux and on the reflector solid angle are
such that comparable parts of the line can come from the absorber and the
reflector. In that case a decrease of $\nh$ would lead to a noticeable decrease
of $\ife$, explaining  the weakness of the \fek\ line in the \asca\/ data.

Figure 4 also shows the 1991--92 spectrum from OSSE.  The OSSE spectrum,
although not simultaneous with other observations presented here, is consistent
with extrapolation of both the \ginga\/ and \asca\/ power laws. Some spectral
steepening at high energies is, however, possible. Fits to the combined
\ginga/OSSE and \asca/OSSE data yield $\ec=1.6_{-1.3}^{+\infty}$ MeV and
$300_{-130}^{+460}$ keV, respectively. The models corresponding to the best-fit
values of $\ec$ are plotted in Figure 4.

\subsection{3C 390.3}
\label{ss:3c390}

\begin{table*} \centering \label{t:3c390a} \caption{Results of fits to the
\ginga, \asca\/ and \exosat\/ spectra of 3C 390.3. The fit to the 1993 November
\asca\/ data set with Compton reflection is limited to the 3--10 keV energy
range. }

 \begin{tabular}{lcccccccccc}
\hline
$A$ & $\as$ & $\nh$ & $\eb$ & $\alpha$ & $\Omega/2\pi$ &
$\efe$ & $\sfe$ & $\ife$ & $\wfe$ &
$\cd(\cnu)$ \\
\multicolumn{11}{c}{\ginga\/ 1988 Nov}\\
$13.3^{+0.2}_{-0.2}$ & -- & $0.0^{+0.2}$ & -- & $0.84^{+0.01}_{-0.01}$ &
0f & $6.33^{+0.31}_{-0.30}$ & 0.1f &
$2.8^{+1.4}_{-1.4}$ & $58$ & 43.8/35(1.25) \\

$14.0_{-0.4}^{+0.5}$ & -- & $0.0^{+0.5}$ & -- & $0.90^{+0.03}_{-0.03}$ &
$0.35^{+0.15}_{-0.24}$ & $6.33^{+0.27}_{-0.27}$ & 0.1f &
$3.4^{+1.3}_{-1.3}$ & $67$ & 28.1/34(0.83) \\
\multicolumn{11}{c}{\ginga\/ 1991 Feb}\\
$5.7^{+0.4}_{-0.2}$ & -- & $0.2^{+1.6}_{-0.2}$ & -- & $0.64^{+0.04}_{-0.03}$ &
0f & $6.36^{+0.41}_{-0.44}$ & 0.1f &
$2.5^{+1.4}_{-1.5}$ &   $84$ & 31.2/35(0.89) \\

$6.2_{-0.7}^{+0.8}$ & -- & $1.6^{+2.3}_{-1.6}$ &  -- & $0.70^{+0.09}_{-0.08}$ &
$0.26^{+0.37}_{-0.26}$ & $6.41^{+0.42}_{-0.46}$ & 0.1f &
$2.7^{+1.4}_{-1.3}$ & 86 & 29.0/34(0.85) \\
\multicolumn{11}{c}{\asca\/ 1993 Nov}\\
 $3.8^{+0.1}_{-0.1}$ & -- & $0.4^{+0.1}_{-0.1}$ & -- & $0.68^{+0.01}_{-0.03}$ &
0f & $6.40^{+0.08}_{-0.07}$ & $0.20^{+0.12}_{-0.13}$ &
$4.6^{+1.0}_{-1.4}$ &   $240$ & 1326/1314(1.01) \\

 $3.9^{+0.1}_{-0.1}$ & $0.74^{+0.04}_{-0.04}$ & $0.5^{+0.1}_{-0.5}$ &
$3.3^{+0.4}_{-0.5}$ & $0.54^{+0.05}_{-0.07}$ & 0f &
$6.38^{+0.07}_{-0.06}$ & $0.10^{+0.12}_{-0.10}$ & $ 3.2^{+1.1}_{-1.1}$ &
$150$ & 1309/1312(1.00) \\

 $3.1^{0.2}_{-0.1}$ & -- & 0.5f & -- & $0.56^{+0.07}_{-0.06}$ & 0.35f &
$6.38^{+0.08}_{-0.06}$ & $0.08^{+0.13}_{-0.08}$ & $2.9^{+1.2}_{-1.0}$ &
$140$ & 545/587(0.93) \\
\multicolumn{11}{c}{\asca\/ 1995 Jan}\\
 $4.5^{+0.2}_{-0.2}$ & -- & $0.8^{+0.2}_{-0.2}$ & -- & $0.70^{+0.04}_{-0.03}$ &
0.35f &$6.44^{+0.11}_{-0.11}$ & $0.00^{+0.33}$ & $2.8^{+1.6}_{-1.4}$ & $120$ &
646/743(0.87) \\
\multicolumn{11}{c}{\asca\/ 1995 May}\\
 $7.8^{+0.4}_{-0.1}$ & -- & $0.4^{+0.2}_{-0.1}$ & -- & $0.77^{+0.03}_{-0.03}$ &
0.35f &$6.75^{+0.39}_{-0.43}$ & $0.20^{+\infty}_{-0.20}$ &
$2.3^{+3.1}_{-1.8}$ & 84 & 1048/1017(1.03) \\
\multicolumn{11}{c}{\exosat\/ 1985--86}\\
5.3  & -- & $0.0^{+4.1}$ & -- & $0.61^{+0.16}_{-0.07}$ &
0.3f & $5.91^{+\infty}_{-5.91}$ & 0.1f & $1.1^{+5.5}_{-1.1}$ &
$32$ & 20/22(0.90) \\
\hline
\end{tabular}
\end{table*}

The optical morphology of this powerful radio galaxy is unusual; there are no
spiral arms, but the luminosity profile does not follow the $r^{1/4}$ law in
the outer parts of the image (Smith \& Heckman 1989). In the radio
frequencies, the object is lobe-dominated and displays superluminal motion
(Wamsteker \& Clavel 1989 and references therein). The galaxy has been
observed by \exosat, \ginga, \asca, and OSSE.

\begin{figure} \centering \epsfxsize=8.3cm \epsfbox{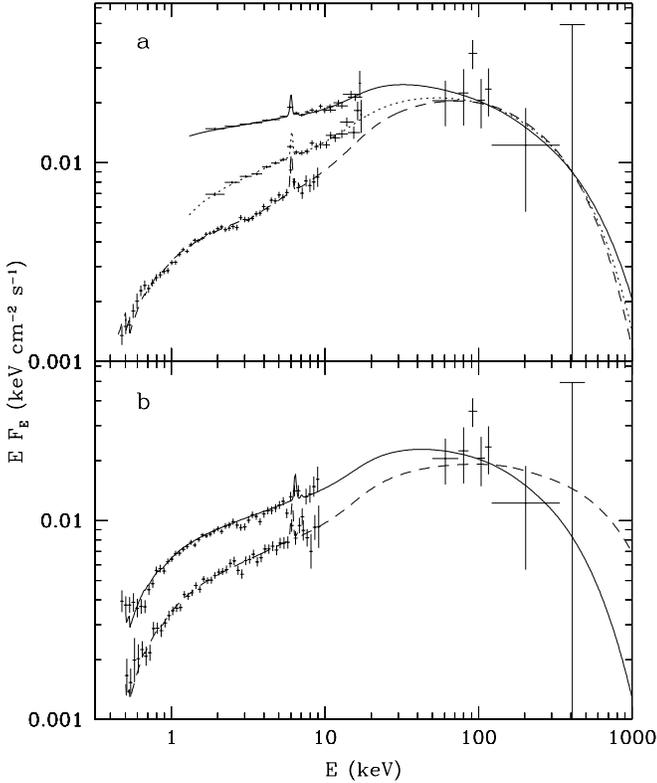}
\label{fig:3c390} \caption{X-ray spectra of 3C 390.3.  {\it (a)} The \ginga\/
spectra from 1988 and 1991, and the \asca\/ spectrum from 1993, from top to
bottom, respectively.  {\it (b)} The \asca\/ spectra from 1995 May (top) and
January (bottom).  Above 50 keV, the average spectrum from 1991--93
observations by OSSE is shown on both panels.  The model curves include
reflection and high-energy cutoffs implied by the OSSE data.  } \end{figure}

We first analyze the spectra from \ginga, shown in Figure 5a. Our 1991 data
exclude a period when a stellar flare occured in the field of view of \ginga\/
(Inda et al.\ 1994). The 1988 observation has been included in the NP94 sample,
and the 1991 observation has been presented by Inda et al.\ (1994), who,
however, have not searched for the presence of Compton reflection. We note a
possible presence of a soft-excess component in the 1988 data, as indicated by
the null value of the best-fit $\nh$ and positive residuals seen in the lowest
\ginga\/ channels (see Fig.\ 3 of Inda et al.\ 1994).

We find a moderate Compton reflection component corresponding to $\Omega/2\pi
\sim 0.3$ in both observations (Table 7), which is similar to the result of
NP94 for the 1988 observation. The presence of reflection is significant in the
(longer) 1988 observation at the 99.99 per cent confidence level. (The
top-layer \ginga\/ data alone imply somewhat stronger reflection, with
$\Omega/2\pi= 0.50_{-0.20}^{+0.26}$ in 1988 and $0.51_{-0.39}^{+0.74}$ in 1991,
which values are still consistent within statistical errors with those in Table
7.)

We have also considered the effect of the reflector ionization. Similarly to
the case of 3C 111, we find that ionization of the reflector has a negligible
effect on $\Omega$; $\xi=0^{+122}$ erg s$^{-1}$ cm$^{-1}$, and $\Omega/2\pi=
0.33^{+0.18}_{-0.17}$ at the upper limit of $\xi$ for the 1988 \ginga\/ data.
Similar results are obtained for the 1991 observation.

Results of \asca\/ observations of 3C 390.3 in 1993 and 1995 have been reported
in Eracleous et al.\ (1996) and Leighly et al.\ (1997), respectively. Here, we
reanalyse those data using the current version of the response and the
effective area of \asca\/ (see Section \ref{s:data}). First, we fit the data
from 1993 with a power-law continuum, a Gaussian line, and neutral absorption,
see Table 7. Our results on both the continuum and the line are similar to
those of Eracleous et al.\ (1996), in particular, $\alpha\simeq 0.7$ and
$\sfe\simeq 0.2$ keV. However, the data show strong evidence for the presence
of a soft excess, as demonstrated by a large reduction of $\chi^2$ (by $-17$
for addition of 2 parameters, which is significant at the $99.98$ per cent
confidence) when the power law is allowed to break. This is consistent with
the possible presence of a soft X-ray excess in the 1988 \ginga\/ data (see
above) as well as in the \exosat\/ data (Ghosh \& Soundararajaperumal 1991).
The broken power law model gives $\alpha \simeq 0.5$--0.6 above $\sim 3$ keV,
and the Fe line both weaker than for the single-power law fit as well as
unresolved at the 1-$\sigma$ level. The energy-width confidence contours and
the line profile are shown in Figures 2 and 3, respectively.

As found above, the \ginga\/ data above show the presence of Compton
reflection. However, including it only weakly affects the results of the
\asca\/ spectral fits, as shown in Table 7. The main effect is that the \fek\
line becomes slightly weaker and narrower than in the case without Compton
reflection. The \asca\/ spectrum modeled including Compton reflection is shown
in Figure 5a.

Results similar to those discussed above are obtained for the two \asca\/
observations of 1995. They are significantly shorter than that in 1993 (see
Table 3), and thus constraints on the spectral parameters are weaker. We find
those data do not require the presence of a spectral break in the continuum,
i.e., the corresponding reductions in $\chi^2$ are statistically insignificant.
The \fek\ line is unresolved at the 1-$\sigma$ level, see Table 7 and Figures 2
and 3. The 1995 \asca\/ spectra are shown in Figure 5b.

The fit results to the average 1985--86 \exosat\/ spectrum are given in the
last row of Table 7. We find this spectrum is very similar to that of the 1991
\ginga\/ observation, therefore we omit it for clarity in Figure 5. Due to the
limited usable energy range, it does not constrain the amplitude of the
Compton reflection, and we give results $\Omega/2\pi=0.3$ only.

\begin{figure} \centering \epsfxsize=8.4cm \epsfbox{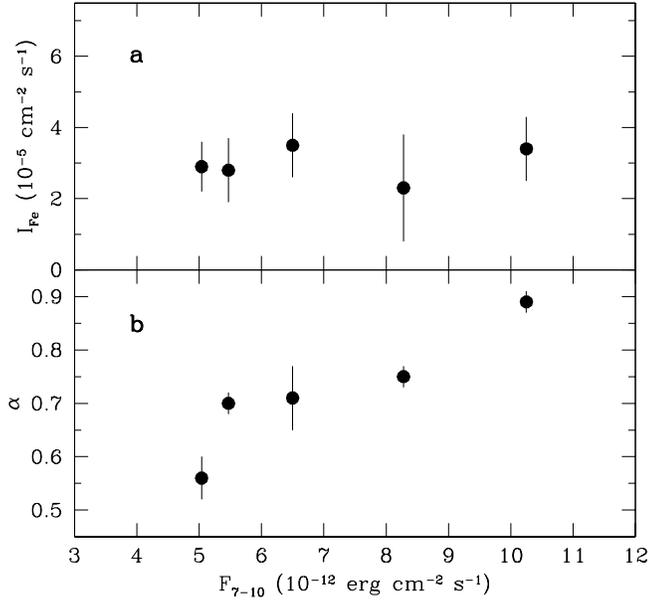}
\label{fig:3c390v} \caption{The dependences of {\it (a)\/} the flux of the
\fek\ line, and {\it (b)\/} the hard X-ray spectral index, $\alpha$, on the
7--10 keV continuum flux in 3C 390.3. The line flux is consistent with being
independent of the continuum level whereas the continuum softens as the source
brightens. The error bars are for the 1-$\sigma$ significance. } \end{figure}

From Table 7, we see that the flux in the Fe line in all observations presented
here is compatible with being constant, $\sim 3\times 10^{-5}$ cm$^{-2}$
s$^{-1}$, whereas the Fe-K ionizing flux ($\ga 7$ keV) varies by a factor of
$\sim 2$, as illustrated in Figure 6a. The correlation coefficient between
$\ife$ and the 7--10 keV flux is $r=0.16$, which corresponds to the probability
that there is indeed no correlation between those quantities of $>45$ per cent.
This argues against the bulk of the line being due Compton reflection from an
inner region of an accretion disk, in which case a proportionality of the line
to the flux would be expected. Furthermore, $\wfe\sim 100$--150 eV in the
observations in 1993 and 1995 January cannot be accounted for by Compton
reflection with $\Omega/2\pi\sim 0.3$ (e.g., GF91), seen in both \ginga\/
observations. (Eracleous et al.\ 1996 claimed the \fek\ line in the 1993
observation could be entirely due to Compton reflection without considering its
stregth measured by \ginga.) On the other hand, the absorbing column is so low
that it can account for at most $\wfe\sim 10$ eV (M86). Possible resolutions of
this issue are discussed in Section \ref{s:dis} below.

Figure 6b shows that the X-ray spectrum softens when it brightens, as
previously noted by Inda et al.\ (1994) and Leighly et al.\ (1997).  This
spectral variability yields spectra pivoting around 100 keV at a flux
corresponding to the average 1991--93 OSSE data, see Figure 5. The OSSE data at
higher energies lie below the extrapolation of the X-ray power laws, which
implies the presence of a high-energy cutoff or break. E.g., the 1988 \ginga\/
data imply $\ec=400_{-170}^{+580}$ keV and the 1993 \asca\/ data yield
$\ec=260_{-110}^{+320}$ keV. Models with the high-energy cutoffs are plotted in
Figure 5.

\subsection{3C 382}
\label{ss:3c382}

\begin{table*} \centering \label{t:3c382} \caption{Results of fits to the
\ginga, \exosat\/ and \asca\/ spectra of 3C 382. The  second fit with
$\ife/\Omega=$ const is for $\af=4$. }

 \begin{tabular}{lcccccccccc}
\hline
$A$ & $\as$ & $\nh$ & $\eb$ & $\alpha$ & $\Omega/2\pi$ &
$\efe$ & $\sfe$ & $\ife$ & $\wfe$ &
$\cd(\cnu)$ \\
\multicolumn{11}{c}{\ginga\/ 1989 Jul}\\
$2.4^{+0.1}_{-0.1}$ & -- & $0.0^{+1.4}$ & -- & $0.50^{+0.06}_{-0.04}$ &
$0.05^{+0.28}_{-0.05}$ & $6.52^{+0.20}_{-0.20}$ & 0.1f &
$3.7^{+1.1}_{-1.2}$ &   $220$ & 29.6/34(0.87) \\
\multicolumn{11}{c}{$\ife/\Omega=$ const} \\
3.0 & -- &  $3.2^{+2.9}_{-2.6}$ & -- & $0.68^{+0.11}_{-0.11}$ &
$0.89^{+0.52}_{-0.89}$ & $6.59^{+0.26}_{-0.26}$ & 0.1f &
$2.5^{+1.5}_{-2.5}$ &   $160$ & 47/35(1.35) \\
2.5 & -- & $0.0^{+2.2}$ & -- & $0.55^{+0.05}_{-0.04}$&
$0.75^{+0.25}_{-0.25}$ & $6.57^{+0.24}_{-0.24}$ & 0.1f &
$3.0^{+1.0}_{-1.0}$ &   $190$ & 35/35(0.99) \\
\multicolumn{11}{c}{\exosat\/ 1983--85} \\
9.6 & -- & $0.0^{+0.3}$ & -- & $0.69^{+0.03}_{-0.02}$ & 0f
& $5.89^{+2.06}_{-0.55}$ & 0.1f & $3.5^{+2.2}_{-2.3}$ &
$72$ & 29.0/21(1.38) \\
9.3 & $0.76^{+0.19}_{-0.05}$ & $0.0^{+2.2}$ & $3.7^{+2.2}_{-1.2}$
&$0.59^{+0.07}_{-0.31}$ & 0f
& $5.35^{+\infty}_{-3.05}$ & 0.1f & $1.7^{+4.6}_{-1.7}$ &
$30$ & 17.8/19(0.94) \\
\multicolumn{11}{c}{\asca\/ 1994 Apr}\\
$14.0^{+0.2}_{-0.3}$ & -- & $0.3^{+0.1}_{-0.2}$ & -- & $1.00^{+0.05}_{-0.04}$ &
0f & $6.57^{+0.91}_{-0.46}$ & $2.34^{+1.17}_{-0.56}$ & $54^{+66}_{-20} $ &
1400 & 1611/1610(1.00) \\

$13.7^{+0.3}_{-0.4}$ & $0.97^{+0.04}_{-0.03}$ & $0.2^{+0.2}_{-0.1}$ &
$3.6^{+0.4}_{-0.5}$ & $0.65^{+0.07}_{-0.07}$ & 0f & 6.30f & 0f &
$1.3^{+1.5}_{-1.3} $ & 27 & 1609/1610(1.00) \\

\hline
\end{tabular}
\end{table*}

The radio morphology of 3C 382 is typical for two-lobe sources with
characteristic hot spots (McDonald, Kenderdine \& Neville 1968). Optically, it
appears as a distorted elliptical (Smith \& Heckman 1989) with extremely broad
Balmer lines reaching FWZI $\sim 25,000$ km s$^{-1}$ (Osterbrock, Koski \&
Phillips 1975).

\ginga\/ observed 3C 382 several times during 5 days in 1989 July (Kaastra,
Kunieda \& Awaki 1991). We have initially analyzed the \ginga\/ spectra from
the observation divided into 3 periods (as in Table 2). They are well described
by a hard power-law spectrum with $\alpha\simeq 0.5$ with neither intrinsic
absorption nor Compton reflection and are almost identical. Therefore, we
consider hereafter only the total spectrum. Table 8 gives the fit results, and
Figure 7 shows the unfolded spectrum and the best-fit model. We see that
allowing for the presence of Compton reflection does not improve the fit, and
$\Omega/2\pi \la 0.3$. (With the top-layer data only, a weaker constraint
is obtained, $\Omega=0.22_{-0.22}^{+0.53}$.) This differs from the
corresponding fit in NP94, who obtained $\Omega/2\pi\ga 1$. However, the
reflection model of NP94 includes the Fe K$\alpha$ line tied to the reflection
continuum (as in our models below) and the strong reflection in their fits is
driven by the large flux in the line. Similarly as for 3C 111 and 3C 390.3,
ionization of the reflector leads to a reduction of the fitted reflection
strength; for a large $\xi=5000$ erg s$^{-1}$ cm$^{-1}$, $\Omega/2\pi
=0^{+0.15}$.

On the other hand, there is a pronounced \fek\ line with $\wfe\simeq 220$ eV in
the data. We find that this line is too strong to be explained by Compton
reflection. Namely, when we tie the strength of the Fe line to the solid angle
of the reflector (as in Section \ref{ss:3c111}), the fit becomes significantly
worse than in the case of the free line flux, even for the Fe abundance 4 times
the solar value (Table 8).

\begin{figure}
\centering
\epsfxsize=8.4cm \epsfbox{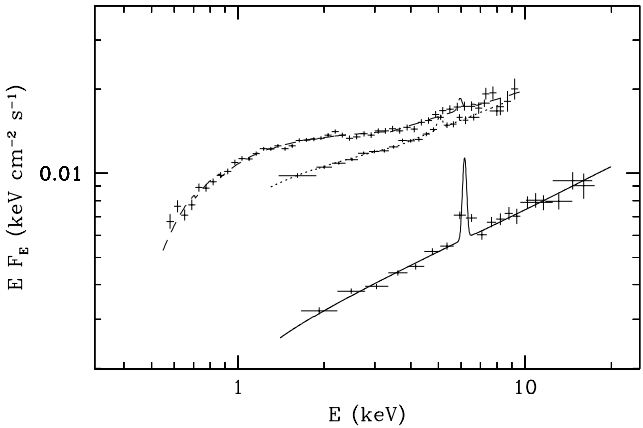}
\label{fig:3c382}
\caption{The X-ray spectra of 3C 382 from \asca, \exosat\/ and \ginga\/ (from
top to bottom). }
 \end{figure}

Kaastra et al.\ (1991) have found the presence of a soft excess in the
\ginga\/ spectrum. We also see a similar component in our data. However, our
analysis is constrained to the range $\geq 1.7$ keV whereas Kaastra et al.\
also included the 1.2--1.7 keV \ginga\/ channel (which calibration is in
general less certain for weak sources). Thus, we do not attempt to constrain
the form of the excess with the \ginga\/ data.

A soft excess is seen very pronouncedly in the \exosat\/ data (Ghosh \&
Soundararajaperumal 1992), which we also find in the average \exosat\/ data,
see Table 8 and Figure 7. Namely, the fit improves with $\Delta \chi^2=-11$
(corresponding to the statistical significance of 99 per cent) when  a break in
the power law spectrum is allowed. The continuum slope above the break becomes
then similar that for the \ginga\/ data.

We then consider the \asca\/ spectrum of 3C 382. We find that adding a soft
excess (using a model with a broken power law) results in a slight improvement
of the fit over the model with a single power-law continuum, as shown in Table
8. However, the two models yield vastly different results regarding the \fek\
line. The single-power law model requires the line to be very strong, $\wfe\ga
1$ keV, and very broad, $\sfe\sim 2$--3 keV (as found before  by Reynolds
1997). The line flux is then $\sim 15$ times larger than that in the \ginga\/
data. Also, the spectral index is much softer than that observed from the
source in all previous \exosat\/ and \ginga\/ observations (Ghosh \&
Soundararajaperumal 1992; Kaastra et al.\ 1991). The extremely large $\sfe$
would require the line origin from reflection in inner parts of an accretion
disk disk (e.g., Fabian et al.\ 1995), for which, however, the expected $\wfe$
is at least several times less. Also, the weakness of the continuum reflection
seen in the \ginga\/ data appears to rule out a reflecting inner disk.

However, the \fek\ line becomes very weak as well as unresolved in the broken
power-law model, and  the hard power-law index becomes than similar to that
observed before by \exosat\/ and \ginga. Also, the line flux in the \asca\/
observation becomes then compatible within statistical errors with that
measured by \ginga. The line parameters are not constrained due to the weakness
of the line, and thus $\efe$ and $\sfe$ have been fixed at the best-fit values
while determining the confidence regions of other parameters in this fit (Table
8). Since the \ginga\/ data show reflection to be very weak, we do show here
results of a fit with Compton reflection. The resulting line profile is shown
in Figure 3.

We interpret the similar $\chi^2$ in both above fits to the \asca\/ data as
simply due to the very broad and strong \fek\ line in the first fit giving the
dominant {\it continuum\/} component above 7 keV, whereas that continuum is
accounted for by the harder power law in the second fit. This discrepancy
illustrates the importance of careful modeling of the continuum for
determination of parameters of \fek\ lines in \asca\/ data, which energy range
extends only slightly above the energy range dominated by the \fek\ line.

We have also analyzed the \asca\/ SIS data of 3C 382 obtained in the bright
mode, which results in a substantially longer exposure for the SIS detectors
but somewhat worse energy resolution (see Section 2). We have found that the
results reported in Table 8 remain virtually unchanged. In particular the
broken power law model yields only a narrow and weak \fek\ line with the
best-fit $\sfe=0$ keV and $\wfe=28$ eV. However, another minimum with almost
the same $\chi^2$ is present, at which the line is broad with $\sfe\simeq 1.4$
keV and $\wfe=230$ eV. This would suggest that reflection from an inner disk
may be present. However, when $\Omega/2\pi$ is fixed at at 1, corresponding to
disk reflection, the line width becomes constrained to $\sfe\la 0.6$ keV with
the best fit at $\sfe\sim 0$ keV and $\wfe\simeq 30$ eV. Thus, the presence of
a strong and broad \fek\ line appears unlikely.

Figure 7 shows that the X-ray continuum of 3C 382 is strongly variable, with
the 2-keV flux changing by an order of magnitude between the \ginga\/ and
\asca\/ observations. The spectral variability is similar to that seen 3C
390.3, namely the spectrum softens with increasing X-ray flux as well as the
\fek\ line flux is, within the statistical errors, compatible with constant.
The line is thus unlikely to be due to reflection from an inner disk, and
probably originates at large distances where the variable continuum is averaged
due to large light-travel time accross the region of line formation. However,
the \fek\ line-emitting region has to be out of the line of sight as the line
strength is much too large (M86) to be due to the weak absorption in this
object.

\subsection{3C 120}
\label{ss:3c120}

3C 120 is a BLRG exhibiting characteristics intermediate between those of FR I
galaxies and BL Lacs (Urry \& Padovani 1995). It is highly variable in all
energy bands, and shows superluminal motion in the VLBI core (Muxlow \&
Wilkinson 1991; Walker, Benson \& Unwin 1988; Walker, Walker \& Benson 1988).

3C 120 has been observed by \exosat\/(Maraschi et al.\ 1991) and \asca, but
not by \ginga. Thus, we haveno reliable informationas yetonthe presence of
Compton reflection in this object. 3C 120was observed by\heao, and the A2
data imply a reflector solid angle of  $\Omega\simeq (1^{+3}_{-1})\times 2\pi$
(rescaling the results of Weaver, Arnaud \& Mushotzky 1995 obtained for
angle-averaged reflection into an inclination of 30$^{\circ}$), i.e., the
measurement errors are large for those data. Also, the A2 results for
reflection in other AGNs (Weaver et al.\ 1995) correspond to solid angles
systematically larger than those from \ginga\/ (NP94). Taking into account
those uncertainties, we compute below the spectral parameters for \asca\/ and
\exosat\/ data for both $\Omega=0$ and $\Omega/2\pi=1$.

\begin{table*}
\centering
\label{t:3c120}
\caption{Results of fits to the \asca\/ and \exosat\/ spectra of 3C
120. The fit with Compton reflection to the \asca\/ data is limited to the
4--10 keV range. }

 \begin{tabular}{lcccccccccc}
\hline
$A$ & $\as$ & $\nh$ & $\eb$ & $\alpha$ & $\Omega/2\pi$ &
$\efe$ & $\sfe$ & $\ife$ & $\wfe$ &
$\cd(\cnu)$ \\
\multicolumn{11}{c}{\asca\/ 1994 Feb}\\
$17.0^{+0.2}_{-0.2}$ & -- & $0.5^{+0.1}_{-0.1}$ & -- & $1.02^{+0.02}_{-0.03}$ &
0f & $6.72^{+0.27}_{-0.23}$ & $1.73^{+0.49}_{-0.41}$ & $40^{+11}_{-10} $ & 1000
& 1393/1441(0.94) \\

$16.7^{+0.3}_{-0.3}$ & $1.00^{+0.02}_{-0.01}$ & $0.4^{+0.1}_{-0.1}$ &
$4.0^{+0.3}_{-0.3}$ & $0.72^{+0.07}_{-0.06}$ & 0f & $6.44^{+0.18}_{-0.22}$ &
$0.32^{+0.76}_{-0.20}$ & $5.3^{+3.0}_{-2.4} $ & 110 & 1383/1439(1.00) \\

$12.7^{+1.3}_{-1.0}$ & -- & 0.4f & -- & $0.82^{+0.06}_{-0.06}$ & 1f &
$6.41^{+0.33}_{-0.21}$ & $0.20^{+\infty}_{-0.20}$ & $3.3^{+6.7}_{-1.7} $ & 63 &
580/617(0.90) \\
\multicolumn{11}{c}{\exosat\/ 1983--86} \\
12.0 & -- & $0.0^{+0.4}$ & -- & $0.79^{+0.02}_{-0.02}$ &
0f & $5.90^{+0.38}_{-0.37}$ & 0.1f & $4.6^{+1.9}_{-2.0}$ &
$91$ & 23/22(1.05) \\

12.9 & -- & $0.4^{+1.0}_{-0.4}$ & -- & $0.88^{+0.04}_{-0.03}$ &
1f & $5.51^{+0.77}_{-0.59}$ & 0.1f & $2.4^{+2.1}_{-2.0}$ &
$41$ & 20/22(0.89) \\
\hline
\end{tabular}
\end{table*}

\begin{figure} \centering \epsfxsize=8.4cm \epsfbox{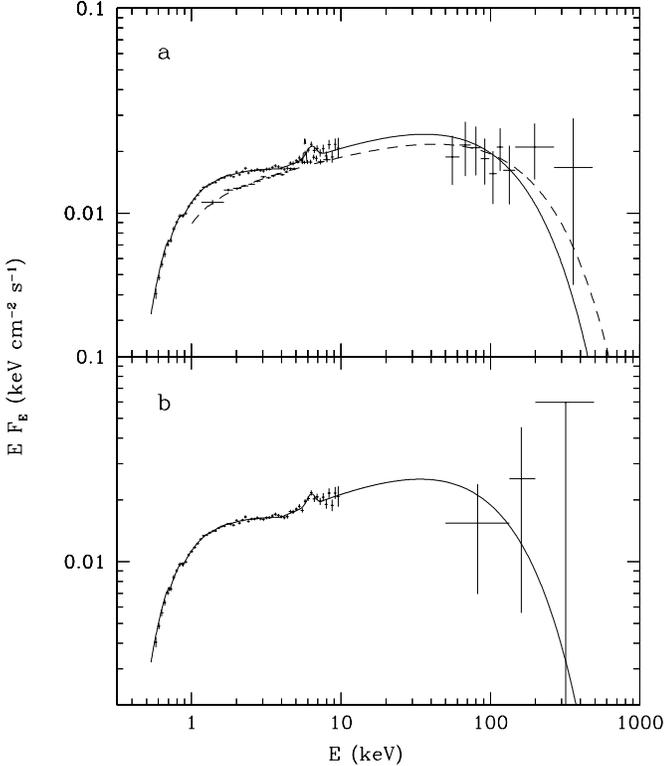}
\label{fig:3c120}

\caption{{\it (a)\/} The spectra of 3C 120 from \asca\/ (the upper X-ray
spectrum), \exosat\/ (the lower X-ray spectrum), and OSSE ($>50$ keV). The
solid and dashed curves correspond to model with a broken power law and an
exponential cutoff fitted to the \asca/OSSE and \exosat/OSSE data,
respectively. {\it (b)} The spectrum of 3C 120 from contemporaneous
observations by \asca\/ (1994 February 17) and OSSE (1994 February 17--March
1). The same model form as in {\it (a)\/} was fitted to the data. }
\end{figure}

Table 9 gives fit results to the \asca\/ spectrum. Similarly to the case of 3C
382, we find the fitted line parameters depend very strongly on the way the
continuum is modeled. If we model the \asca\/ continuum as a single power law,
the line is very strong, $\wfe\simeq 1$ keV, and broad, $\sfe\simeq 2$ keV (as
obtained before by Reynolds 1997). On the other hand, the line strength becomes
moderate, $\wfe\simeq 100$ eV, and relatively narrow, when the power-law
continuum is allowed to break. The spectrum corresponding to the latter model
is plotted in Figure 8a. Allowing a break in the power law also improves the
fit at the significance of 99.6 per cent. Thus, the very strong and broad line
obtained for a single power-law fit seems to be an artefact of underestimating
the continuum around the line in this model.

The presence of a soft excess in 3C 120 is confirmed by Grandi et al.\ (1997),
who analyzed \rosat\/ PSPC observations from 1993. Those data, when fitted by a
power law, show $\alpha \sim 2$--3 in the 0.15--2.1 keV range. This argues in
favour of the broken power-law model for the \asca\/ data and against the
reality of the very strong and broad \fek\ line obtained assuming no soft
excess. Also, the averaged \exosat\/ data yield almost the same parameters for
the broken-power law model as the \asca\/ data, although those parameters are
only weakly constrained.

The line parameters depend somewhat on the presence of Compton reflection.
Table 9 also presents results of a fit with the assumed $\Omega/2\pi=1$ to the
4--10 keV data. The line becomes then even weaker and narrower than for the
broken-power law model without reflection. The confidence contours for $\efe$
and $\sfe$ and the line profile are shown in Figures 2 and 3, respectively.
Note that this fit yields a small $\wfe\simeq 60$ eV, which is not compatible
with theoretical predictions (GF91; \.Zycki and Czerny 1994) for the line
strength from this process of $\wfe\sim 150$--200 eV (unless the reflector is
so strongly ionized that resonant absorption suppresses the line), which argues
against $\Omega/2\pi\ga 1$ in 3C 120.

The \asca\/ spectrum of 3C 120 has also been fitted in the 3--10 keV range
(without including reflection) by Nandra et al.\ (1997). They obtained a
Gaussian line with $\sfe= 0.74_{-0.27}^{+0.34}$ keV and $\wfe= 330_{-
130}^{+200}$ eV, and the power law index of $\alpha= 0.89_{-0.06}^{+0.08}$
(their confidence regions are for $\Delta\chi^2=4.7$). To enable a precise
comparison of the results, we have also performed a fit with a power-law model
in the 3--10 keV range, and obtained results virtually identical to those for
the broken power-law fit presented in Table 9, e.g., $\sfe=0.28^{+0.96}_{-
0.16}$ keV and $\wfe\simeq 100$ eV. Thus, we find the \fek\ line to be
significantly narrower and weaker than that found by Nandra et al.\ (1997) for
the same model and the same energy range used. This difference is apparently
caused by an older version of the \asca\/ response and effective area used by
those authors (of 1995, whereas we use the release of 1997). To study the issue
further, we have also used the \asca\/ data on 3C 120 obtained with the data
processing package of 1996 March-April. We have found very little difference
with respect to our current results, e.g., $\sfe=0.26^{+0.76}_{- 0.17}$ keV and
$\wfe\simeq 90$ eV for the power-law model in the 3--10 keV range. This
demonstrates the importance of updating the results on the \fek\/ line
parameters obtained with the pre-1996 data-processing software.

This conclusion is further reinforced by a comparison of our results with those
of Grandi et al.\ (1997). They have fitted the 0.6--10 keV \asca\/ spectrum
with a broken power-law continuum model. They found the break energy of 4 keV,
similarly to our results (Table 9). However, they obtained the parameters very
similar to those of Nandra et al.\ (1997), $\sfe= 0.71_{- 0.25}^{+0.24}$ keV
(in the source frame), $\wfe= 280_{-80}^{+160}$ eV, and $\alpha=
0.88_{-0.08}^{+0.10}$, i.e., a line much broader and stronger and the hard
continuum softer than those found for the same model in Table 9. These
differences can again be entirely explained by an earlier version of the
processing software (of 1994) they used.

The \asca\/ observation was contemporaneous with an observation by OSSE data
(VP 317, see Table 4). This allows us to study the presence of a high-energy
cutoff in the spectrum. We used a continuum model consisting of a  broken power
law multiplied it by an exponential factor. The resulting X\g\ spectrum is
shown in Figure 8b. We see that a high-energy spectral break or cutoff is
indeed present, with the an e-folding energy of $\ec=110^{+130}_{-50}$ keV. The
fit with no cutoff yields $\Delta\chi^2 =+12$, which corresponds to the
statistical significance of the presence of a cutoff (or a break) of 99.95 per
cent. We have also used the sum of the OSSE data for VP 317 and 320 (see Table
4), and found similar results. The significance of the presence of a cutoff
increases if Compton reflection is assumed to be present (although the best fit
to the combined data is obtained at $\Omega=0$). Thus, the cutoff cannot be
explained as an artefact of the spectral curvature introduced by the spectral
component from Compton reflection. We note that our result differs from that of
Grandi et al.\ (1997), who found no cutoff in the combined data. This
difference is due to, first, the \asca\/ data best-fitted by them by a softer
power law than ours (due to an older version of the processing software, see
above), and, second, their 50--150 keV flux from OSSE higher by $\sim 30$ per
cent than that found by us (which difference we have not been able to explain).

We can further constrain the presence of a cutoff by using the average OSSE
spectrum from 9 observations in 1992--97 (see Table 4), shown in Figure 8a. If
we fit those data together with the \asca\/ data, we obtain $\ec=
130_{-40}^{+150}$ keV. We also compare the average OSSE spectrum with the
average \exosat\/ spectrum from 14 observations in 1983--86, also shown in
Figure 8a. The joint fit with an e-folded power-law model yields
$\ec=200_{-60}^{+100}$ keV, confirming the presence of a spectral break between
X-rays and soft \g-rays.

\section{OVERALL PROPERTIES OF BROAD-LINE RADIO GALAXIES}
\label{s:average}

\subsection{Summary of individual properties}
\label{ss:x_i}

\begin{table*} \centering \label{t:summary} \caption{Summary of main parameters
of BLRGs. The statistical sifnificance of the presence of an \fek\ line is
given by $P$, where $1-P$ given below is the probability that there is no line
in the data. The e-folding energy, $\ec$, is obtained using the average OSSE
data when the data exist and assumed to be 400 keV otherwise, and $L$ is the
absorption-corrected 1--1000 keV isotropic luminosity corresponding to that
model in units of $10^{44}$ erg s$^{-1}$ (assuming $H_0=75$ km s$^{-1}$
Mpc$^{-1}$). }

 \begin{tabular}{lcccccccc} \hline Object & Instrument & $\alpha$ &
$\Omega/2\pi$ & $\wfe$ & $\sfe$ & $1-P$ & $\ec$ [keV] & $L$ \\

3C 445 & \ginga & $0.45^{+0.14}_{-0.10}$ & $0^{+0.22}$ & 150 & 0.1f & -- & 400f
& 13.5 \\

& \asca & $0.39^{+0.28}_{-0.29}$ & 0f & 140 & $0^{+0.18}$ & $1.4\times 10^{-3}$
&
400f & 29.4 \\

3C 111 & \ginga & $0.82^{+0.08}_{-0.06}$ & $0.12^{+0.31}_{-0.12}$ & 78 & 0.1f &
0.037 & $1600^{+\infty}_{-1300}$ & 10.0 \\

& \asca & $0.73^{+0.03}_{-0.02}$ & 0.12f & 25 & $0^{+\infty}$  & 0.73 &
$300^{+460}_{-130}$ & 9.2 \\

3C 390.3 & \ginga & $0.90^{+0.03}_{-0.03}$ & $0.35_{-0.24}^{+0.15}$ & 67 & 0.1f
& $4.7\times 10^{-4}$ & $400^{+580}_{-170}$ & 11.6 \\

& & $0.70^{+0.09}_{-0.08}$ & $0.26_{-0.26}^{+0.37}$ & 86 & 0.1f &
0.013 & $300^{+420}_{-120}$ & 9.0 \\

& \asca & $0.54^{+0.05}_{-0.07}$ & 0f  & 150 & $0.10_{-0.10}^{+0.12}$ &
$5.6\times 10^{-8}$ & $260^{+320}_{-110}$ & 7.8 \\

& & $0.70^{+0.04}_{-0.03}$ & 0.35f & 120 & $0^{+0.33}$ & 0.016 &
$620_{-350}^{+3600}$ & 8.7 \\

&& $0.77^{+0.03}_{-0.03}$ & 0.35f & 84 & $0.20^{+\infty}_{-0.20}$ &
0.28 & $300_{-110}^{+300}$ & 9.8 \\

3C 382 & \ginga & $0.50^{+0.06}_{-0.04}$ & $0.05_{-0.05}^{+0.28}$ & 220 & 0.1f
& $1.0\times 10^{-5}$ & 400f & 8.4 \\

& \asca & $0.65^{+0.07}_{-0.07}$ & 0f & 25 & $0^{+\infty}$ & 0.57 & 400f & 17.2
\\

3C 120 & \asca & $0.72^{+0.07}_{-0.06}$ & 0f & 110 & $0.32_{-0.20}^{+0.76}$ &
$3.0\times 10^{-5}$ & $130_{-40}^{+150}$ & 3.8 \\

\hline
\end{tabular}
\end{table*}

Table 10 presents main parameters of selected models fitted to the X-ray
spectra in Section 4, and the significance of the presence of the \fek\ line
obtained using the $F$-test. It also gives the e-folding energies obtained
using the available OSSE data, and the corresponding X\g\ intrinsic
luminosities. We see that the continuum Compton reflection is either absent or
weak but the \fek\ line flux is often much stronger than that expected from the
weak reflection. Furthermore, this is the case in both objects with weak
absorption in the line of sight and in 3C 445, which has strong ($\nh\ga
10^{23}$ cm$^{-2}$) absorption.

In 3C 445, the \fek\ line can be explained entirely by absorption, and Compton
reflection is constrained to be weak, $\Omega/2\pi \la 0.2$. In 3C 111,
$\Omega/2\pi \la 0.4$, and the flux in the \fek\ line can be in principle
explained by the sum of the contributions from the observed absorption and
Compton reflection. The only object with Compton reflection detected
unambiguously is 3C 390.3, in which $\Omega\sim 0.3$ was detected with a 99.99
per cent significance in one \ginga\/ observation. However, the \fek\ line
strengths in two \asca\/ observations of this object were too large to be
explained by that reflection (as well as absorption was negligible). In 3C 382,
$\Omega/2\pi \la 0.3$ was obtained from the \ginga\/ observation, but the line
flux in that observation was much too large to be explained by either
reflection or (negligible) absorption. In 3C 120, Compton reflection has not
been measured. However, $\Omega/2\pi\sim 1$ appears unlikely based on the
weakness of the \fek\ line implied by a model with strong reflection.

The observed variability also indicates the origin of the \fek\ line different
than from Compton reflection from an accretion disk. Namely, the line flux in
repeated observations of 3C 390.3 and 3C 382 over several years by \ginga\/ and
\asca\/ is compatible with being constant whereas the flux around 10 keV
changes within a factor of several. This disagrees with the origin of the line
from the hard X-ray continuum incident on an accretion disk.

Furthermore, all but one \asca\/ observations show the lines are consistent
with being unresolved with respect to the continua fitted above $\sim 3$ keV.
Thus, these data provide no compelling argument for the origin of the lines in
inner accretion disks, consistent with the weakness of reflection and the lack
of line variability. A possible exception is 3C 120, in which the broken-power
law model without reflection yields the line being moderately broad, see Table
10. We note that if the 0.5--10 keV continuum in 3C 382 and 3C 120 is forced to
be a single power law, the fitted lines become resolved and much stronger, with
$\sfe\sim 2$ keV and $\wfe\sim 1$ keV. However, this is appears to be an
artefact of neglecting the soft X-ray excesses present in BLRGs (observed
by, e.g., \exosat\/ and \rosat).

Table 10 also shows results of fitting the X-ray data of 3 objects together
with the OSSE data (not simultaneous) with an e-folded power law continuum. We
note that in {\it all\/} cases, the OSSE data lie on or below the extrapolated
X-ray continuum (initially fitted without the OSSE data), in spite of strong
variability of both the X-ray flux and spectral index. This argues strongly for
the presence of soft \g-ray spectral breaks in those spectra.

\subsection{The average X-ray properties}
\label{ss:x}

The average hard X-ray spectral index of the studied objects is $\sim 0.7$ for
observations by all the three considered instruments, \ginga, \asca, and
\exosat.\/ When Compton reflection is included, the average $\alpha$ from the
\ginga\/ observations is 0.67 with a dispersion of 0.18. Similar values are
obtained for the \exosat\/ and \asca\/ observations. Only the \ginga\/ data
constrain the relative strength of Compton reflection. The unweighted average
for the 5 observations of the 4 BLRGs is $\Omega/2\pi= 0.16$.

\begin{figure}
\label{fig:average}
\centering \epsfxsize=8.4cm
\epsfbox{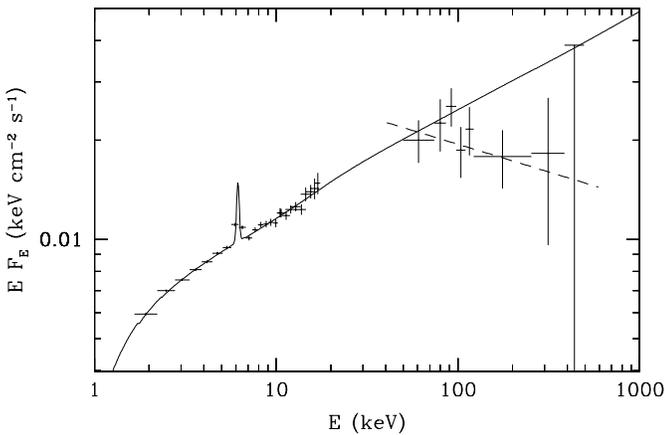}
\caption{The average \ginga\/ spectrum of 3C 111, 3C 382, and 3C 390.3, and
the average OSSE spectrum of 3C 111, 3C 120, and 3C 390.3. The \ginga\/
spectrum is fitted by the model with a power-law incident continuum plus
Compton reflection and the \fek\ line (solid curve). The dashed curve gives
the power-law fit to the OSSE spectrum. }
 \end{figure}

Very similar results to those above are also obtained for the average \ginga\/
spectrum of BLRGs, see Figure 9 and Table 11 (excluding 3C 445 due to the
contamination of its spectrum by a cluster as well as due to its strong
intrinsic absorption). The average values of $z$ and $N_{\rm H,G}$ for the
sample have been used. If only the top-layer \ginga\/ data are used,
$\Omega=0.09_{-0.09}^{+0.25}$ is found, i.e.,  very similar to that given in
Table 11.

\begin{table*}
\centering
\label{t:average}
\caption{Fits to the average \ginga, OSSE and \exosat\/ spectra of BLRGs.}

 \begin{tabular}{lccccccccc}
\hline
$A$ & $\alpha$ & $N_{\rm H}$ & $E_{\rm b}$ & $\Omega/2\pi$ &
$\alpha_{\rm h}$ & $\efe$ & $\ife$ & $\wfe$ &
$\cd(\cnu)$ \\
\multicolumn{10}{c}{Average \ginga}\\
$5.5$ & $0.67^{+0.05}_{-0.04}$ & $1.5^{+1.0}_{-1.1}$&
-- & $0.08_{-0.08}^{+0.17}$ & -- & $6.49^{+0.13}_{-0.14}$&
$3.4^{+0.7}_{-0.7}$&
130 &35/34(1.02) \\
\multicolumn{10}{c}{Average OSSE}\\
46.7 & -- & -- & -- & 0f  & $1.15^{+0.25}_{-0.28}$ & -- & -- &
-- &35/38(0.97) \\
\multicolumn{10}{c}{\ginga/OSSE}\\
9.2 &$0.76^{+0.02}_{-0.02}$ & $2.5^{+0.7}_{-0.7}$&
$95^{+20}_{-25}$& 0f & $2.14^{+3.10}_{-0.90}$& $6.46^{+0.18}_{-0.17}$&
$3.1^{+0.9}_{-0.8}$&
80 &50/78(0.65) \\
\multicolumn{10}{c}{\exosat/OSSE}\\
10.6 &$0.77^{+0.02}_{-0.02}$& 0f &
$80^{+35}_{-73}$ & 0f &  $1.52^{+1.20}_{-0.55}$& $5.9^{+0.4}_{-0.5}$&
$3.9^{+1.9}_{-1.9}$& 76 &49/54(0.90) \\
\multicolumn{10}{c}{Average \ginga/average OSSE}\\
5.4 &$0.67^{+0.02}_{-0.02}$ & $1.2^{+1.0}_{-0.7}$& $93_{-28}^{+20}$ & 0f &
$1.66_{-0.55}^{+1.02}$ & $6.47^{+0.14}_{-0.13}$&
$3.4^{+0.8}_{-0.7}$&
130 &66/70(0.94) \\

\hline
\end{tabular}
\end{table*}

We stress that the average continuum properties of our sample are thus
distinctly different from those of the sample of Seyfert 1s of NP94, which is
strongly dominated by radio-quiet objects (although it also includes 3C 111, 3C
382, and 3C 390.3). For those objects, the average $\alpha=0.95$ with a
dispersion of 0.15 and the average solid angle of the Compton reflector (in
fits with a \fek\ line independent of the continuum reflection) is $\Omega/2\pi
=0.52$ with a dispersion of 0.18. So the two samples are distinctly different
statistically.

In spite of the weak reflection in BLRGs, the \ginga\/ and \asca\/ data show in
many cases strong \fek\ lines in the spectra, with the average equivalent width
of 100 eV. This equivalent width clearly cannot be explained by the weak
Compton reflection observed (e.g., GF91).

\subsection{The soft $\bmath{\gamma}$-ray properties}
\label{ss:gamma}

OSSE has observed 3 BLRGs until 1997, see Table 4. Table 4 also gives the
values of the spectral index for fits with a power law, which are plotted in
Figure 10. Although the individual determinations of $\alpha$ bear large
statistical errors, both the average $\alpha=1.17\pm 0.16$, and the $\alpha$
fitted to the average OSSE spectrum, see Table 11 and Figure 9, are
significantly larger than the X-ray spectral indices. Since Compton reflection
is weak in BLRGs, the difference between the spectral indices clearly shows the
existence of a break around $\sim 100$ keV in the intrinsic soft \g-ray
spectrum, see Figure 9.

Interestingly, the average OSSE spectrum of BLRGs is itself fully consistent
with a power law and does not show any evidence for curvature. When the power
law model is replaced by an e-folded power law, there is no improvement of the
fit ($\Delta\chi^2=-1$ for addition of one parameter). Thus, there is no
evidence for a high-energy cutoff in the power law above $\sim 100$ keV (apart
from the spectral break at that energy implied by the extrapolated X-ray
spectra). This contrasts the case of radio-quiet Seyferts 1s and 2s, which
average OSSE spectra show spectral curvature at very high significance:
$\chi^2_\nu=1.4$ for a power law fit and 0.9 for an e-folded power law fit
(Johnson et al.\ 1997).

\begin{figure} \label{fig:osse_index} \centering \epsfxsize=8.4cm
\epsfbox{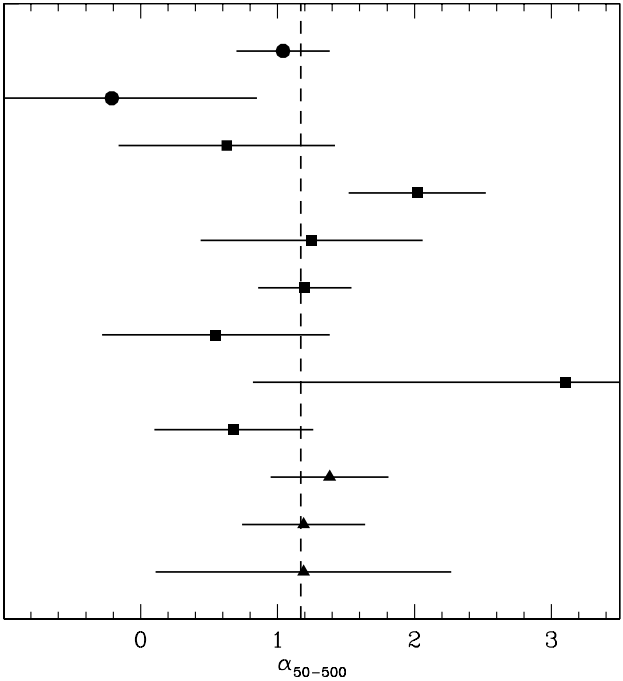} \caption{The distribution of the 50--500 keV energy
spectral index in OSSE observations of BLRGs. The vertical dashed line
corresponds to the weighted average of $\alpha$. The filled circles, squares
and triangles with error bars correspond to the fits in Table 4 for 3C 111, 3C
120, and 3C 390.3, respectively. } \end{figure}

\begin{figure}
\label{fig:broken}
\centering
\epsfxsize=8.4cm \epsfbox{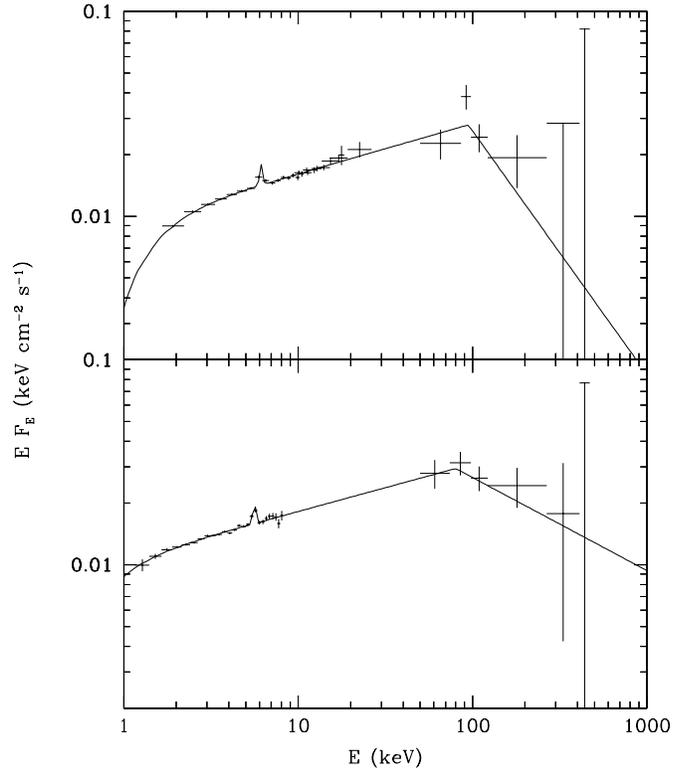}
\caption{The average spectrum of 3C 111 and 3C 390.3 from \ginga/OSSE (top)
and of 3C 120 and 3C 390.3 from \exosat/OSSE (bottom). The solid curves
correspond to the model with a broken power law and an \fek\ line. }
 \end{figure}

A break at $\sim 100$ keV is also indicated by the simultaneous \asca/OSSE
observation of 3C 120 (see Section \ref{ss:3c120}). To further constrain the
presence of a break, we have also obtained the average X\g\ spectra of BLRGs
observed by both \ginga\/ and OSSE (3C 111 and 3C 390.3; Z95), and \exosat\/
and OSSE (3C 120 and 3C 390.3). In fits, we use the average values of $z$ and
$N_{\rm H,G}$ from Table 1 for each pair. Since the number of OSSE observations
is limited for those samples, we allow for free normalization of the OSSE data
with respect to the other data sets.

We have fitted the spectra with e-folded power laws and Compton reflection, and
with broken power laws, and found that the latter model provides a better fit
($\Delta \chi^2=- 6$ and $-2$ for \ginga/OSSE and \exosat/OSSE, respectively).
Results of the fits with that model are are given in Table 11 and Figure 11
(the \exosat/OSSE data are best-fitted with $\nh=0$). The fits confirm the
presence of breaks in those spectra around $\sim 100$ keV although the form of
the break is relatively poorly constrained. A single power-law model is ruled
out at a high significance, $\Delta\chi^2=+10$ and $+6$ for the \ginga/OSSE and
\exosat/OSSE data, respectively.

Finally, we also fit together the average \ginga\/ and OSSE spectra of all
available objects, shown in Figure 9, see the last row of Table 11. Again an
e-folded power law model provides a worse fit than a broken power law, and the
break energy is at $\sim 100$ keV.

\section{Comparison with C{\lowercase{en}} A}
\label{s:cena}

\begin{table*} \centering \caption{Results of the fit to the \ginga\/ spectrum
of Cen A. The model also includes 3 fixed components, see text. $\wfe$ is given
with respect to the total absorbed continuum. }

\begin{tabular}{ccccccccc}
\hline
$A_i$ & $N_{{\rm H},i}$ & $\alpha$ & $\Omega/2\pi$ & $\efe$ & $\sfe$ & $\ife$ &
$\wfe$ & $\cd(\cnu)$ \\

$120_{-10}^{+10}$, $4.5_{-1.6}^{+3.5}$ & $170^{+10}_{-10}$,
$10.0^{+1.2}_{-1.0}$  & $0.79^{+0.06}_{-0.05}$ & $0^{+0.15}$ &
$6.49^{+0.11}_{-0.11}$ &0.1f & $42^{+8}_{-9}$ & $130_{-25}^{+25}$ & 40/39(1.03)
\\

\hline \end{tabular} \end{table*}

 \begin{figure}
\label{fig:cen_a}
\centering
\epsfxsize=8.4cm \epsfbox{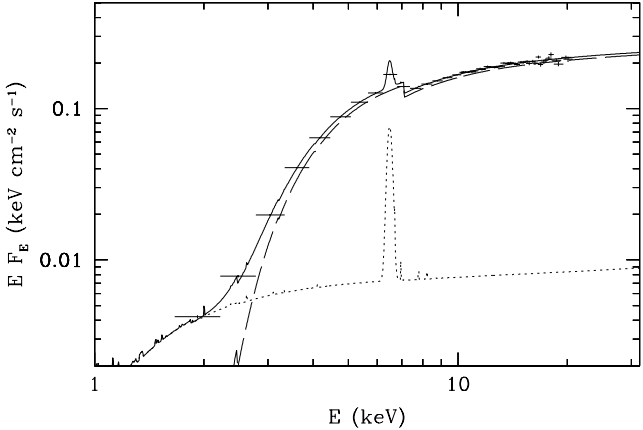}
\caption{The \ginga\/ spectrum of Cen A (crosses). The dashed curve represents
the absorbed hard X-ray power law, and the dotted curve gives the sum of
components dominant at soft X-rays (see text) and the \fek\ line. The solid
curve gives the sum. No Compton reflection is found in the best fit of this
model.
 }
 \end{figure}

Cen A (NGC 5128) is the nearest active galaxy at $z=0.0008$ and with $N_{\rm
H,G} =7\times 10^{20}$ cm$^{-2}$ (Stark et al.\ 1992). No broad line emission
can be observed from this giant elliptical radio galaxy due to heavy absorption
in a warped dusted lane viewed close to edge-on, and thus it is not a BLRG. On
the other hand, the unified AGN model (e.g., Antonucci 1993) postulates that
the difference between AGNs with and without broad lines is solely due to the
viewing angle of the nucleus. Since Cen A is one of the the brightest
extragalactic X-ray sources, we can determine with high accuracy its X\g\
properties despite the relatively large absorbing column density, in
particular, the presence of Compton reflection and a high-energy break. We can
then compare these properties with those obtained in this work for BLRGs, with
the goal to test the unified AGN model.

We use here the \ginga\/ spectrum of Cen A of 1989 March 8 (Miyazaki et al.\
1996; Warwick et al.\ 1998). We note that Cen A exhibits a strong and complex
soft X-ray spectrum. To account for that, we include fixed soft X-ray
components due to the jet and diffuse emission (observed by {\it ROSAT}) as a
soft power law and 2 hot-plasma components following Turner et al.\ (1997). The
primary nuclear continuum is modeled by a power law including reflection
absorbed by a column, $N_{{\rm H},1}$. The viewing angle of the reflection
component is taken as $70\degr$ (Graham 1979; Dufour et al.\ 1979). The
nucleus continuum itself shows a soft X-ray excess, which we model here as
a power law with the same $\alpha$ but absorbed by a lower column, $N_{{\rm
H},2}$, following Warwick et al.\ (1998) and Turner et al.\ (1997). This
provides a good fit to the \ginga\/ spectrum, and fit results are presented in
Table 12. We see that Compton reflection is at most weak, similar to results of
Miyazaki et al.\ (1996) and Warwick et al.\ (1998). (If only the top-layer data
are used, $\Omega/2\pi=0^{+0.15}$, identical to the result in Table 12.) The
\fek\ line in the spectrum can be explained as due to fluorescence in the
absorbing column, $\nh$ (M86). Note that when the Fe K$\beta$/Ni K$\alpha$
lines are included in the model (as for the \asca\/ data, Section 3),
$\efe\simeq 6.4\pm 0.1$ keV and $\sfe\simeq 0^{+0.4}$ keV. The line is thus
compatible with being narrow, as confirmed using \asca\/ data by Turner et al.\
(1997). We also note that our results on the hard continuum are rather
independent of the assumed form of the soft X-ray excess, which is only weakly
constrained by the \ginga\/ data. For example, we have obtained almost
identical results as above by modeling the entire soft component as
bremsstrahlung at $kT=1.65$ keV, which was used in fitting \exosat\/ data on
Cen A by Morini, Anselmo \& Molteni (1989).

OSSE observed repeatedly Cen A during 1991--94 (Kinzer et al.\ 1995). The soft
\g-ray flux varied within a factor of two. All the spectra showed distinct
spectral breaks. When the spectra were fitted with a broken power law, the
break energy was $\sim 140$--170 keV. The power law below the break was
$\alpha\sim 0.7$ (only slightly harder than that seen by \ginga\/ in 1991). A
similar break energy of 180 keV was found in a balloon observation (Miyazaki et
al.\ 1996). The hard power law seen by OSSE above the break was found to
continue without any further break to $\sim 10$ MeV by the COMPTEL detector
aboard {\it CGRO\/} (Steinle et al.\ (1997), as illustrated in Fig.\ 6 of
Johnson et al.\ (1997). However, a spectral softening above $\sim 10$ MeV is
required by the $\ge 100$ MeV flux measured from Cen A by EGRET (see Steinle et
al. \ 1997).

Thus, we find that the X-ray spectral index, the tight limit on Compton
reflection, and the spectral break around $\sim 150$ keV in Cen A are very
similar to those found in our sample of BLRGs. (Unfortunately, no useful
constraints on the spectra above 1 MeV can be obtained for BLRGs). This
similarity is strongly suggestive of the common nature of the radiative
processes in both narrow-line and broad-line radio galaxies, as expected in the
AGN unified model.

\section{DISCUSSION} \label{s:dis}

\subsection{The origin of the Fe K$\bmath{\alpha}$ line}
\label{ss:line}

Our results clearly show that the bulk of the \fek\ line flux in our sample of
BLRGs is unlikely to be due to Compton reflection, which is weak (or absent)
in the studied objects. In 3C 445, the line can be satisfactorily explained by
strong absorption in the line of sight, with $\nh\ga 10^{23}$ cm$^{-2}$. This
is also the case in our comparison narrow-line object, Cen A. On the other
hand, absorption in the line-of-sight in 3C 111, 3C 120, 3C 382, and 3C 390.3
is too weak to be able to explain the observed line fluxes.

However, the observed line fluxes can be satisfactorily explained in those
objects if there is matter with $\nh\ga 10^{23}$ cm$^{-2}$ covering a large
solid angle, $\Omega_{\rm o}$, {\it outside\/} our line of sight. Such matter
can be identified with molecular torii, common in AGNs. Irradiation of this
matter by a power law spectrum with the 1-keV normalization, $A$, and the
energy spectral index, $\alpha$, gives rise to a flux in the \fek\ line of
approximately
\begin{equation} \label{eq:line} \ife \approx {1.7\af \over
\alpha+3} {\Omega_{\rm o}\over 4\pi} Y A
\min\left(\taut,{0.5\over \af} \right) {(1\,{\rm keV})^{1+\alpha} \over E_{\rm
K}^{\alpha}},
\end{equation}
where $1.7\af$ is the ratio of the Fe K-edge cross section (at
$E_{\rm K}$) times the relative Fe abundance to the Thomson cross section,
$Y\sim 1/2$ is the fluorescence yield, and $\taut$ is the Thomson optical depth
of the medium (see Krolik \& Kallman 1987). At $\taut\ga 0.5/\af$, the K-edge
optical depth exceeds unity and the line flux approximately saturates. For the
average \ginga\/ spectrum of BLRGs (Table 11), $(\Omega_{\rm o} /4\pi)\af
\taut\simeq 0.1$ is required to account for the observed \fek\ line. This is
compatible with the $\nh$ observed in 3C 445 and Cen A.

\begin{figure} \label{fig:scat} \centering \epsfxsize=8.4cm
\epsfbox{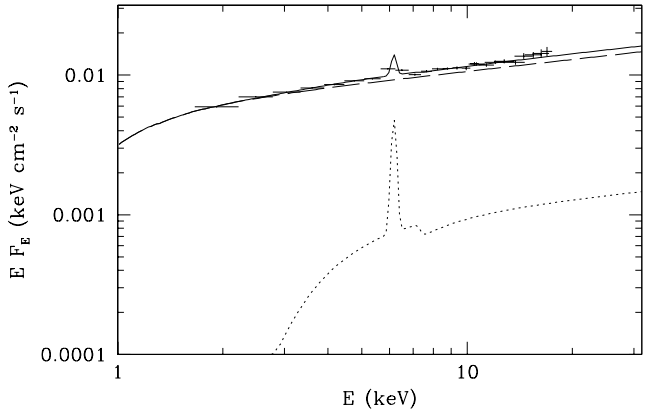} \caption{A contribution to the spectrum of a BLRG (as
illustrated by the average \ginga\/ spectrum) from Thomson scattering in a
medium with $\nh=2\times 10^{23}$ cm$^{-2}$ covering a $3\pi$ solid angle as
seen from the nucleus (dotted curve). The medium is outside of the line of
sight and thus it does not absorb the direct continuum with assumed no
reflection (dashed curve). The solid curve gives the sum. } \end{figure}

Such a medium will also give rise to a scattered continuum component,
approximately given by $(\Omega_{\rm o}/4\pi)\taut \exp[-\tau_{\rm bf}(E)]$
times the observed direct continuum, where $\tau_{\rm bf}(E)$ is the bound-free
optical depth of the surrounding medium. The presence of such a continuum
component is allowed by the X-ray data (but only weakly constrained). Figure 13
shows an example of the scattered spectrum (the dotted curve) for $\nh=2\times
10^{23}$ cm$^{-2}$ and $\Omega_{\rm o}\simeq 3\pi$. We see that although this
component contributes only weakly to the total continuum it accounts entirely
for the \fek\ line. We note that the scattered component does not have the
shape characteristic to Compton reflection (with a fast increase of its
relative contribution around 10 keV). Thus, the weak Compton reflection still
present in the X-ray spectra of BLRGs cannot be explained by scattering in the
surrounding optically-thin matter (see Section \ref{ss:refl} below).

We note that the surrounding medium will be, most likely, partly photo-ionized.
For an ionizing luminosity typical for a BLRG in our sample, $L_{\rm ion} \sim
5\times 10^{44}$ erg s$^{-1}$, and a distance between the nucleus and the
medium with $\nh\sim 2\times 10^{23}$ cm$^{-2}$ of 5 pc, the ionization
parameter (see Section 3) will be $\xi\sim 100$. Therefore, we assumed
$\xi=100$ in the illustrative example in Figure 13. At a typical medium
temperature of $10^4$ K, the dominant Fe ion is found to be Fe {\sc xv},
implying that resonant absorption of the line is negligible. We note that a
range of ionization states in the line-producing medium will give rise to some
broadening of the line.

A possible complication in the scenario with production of \fek\ photons
outside of the line of sight can be due to beaming of the primary radiation
(e.g., Awaki et al.\ 1991). Beaming is expected in jet emission, and radio
jets are present in most of the sources in our sample. Such beaming will relax
the constraint on $(\Omega_{\rm o}/4\pi)\taut$ discussed above. For example,
we could explain the line in 3C 111 seen by \ginga\/ as due to fluorescence in
a medium with the {\it observed\/} $\nh$ ($\sim 2\times 10^{22}$ cm$^{-2}$) if
the irradiating flux is enhanced by a factor of $\sim 100$ in a cone with the
opening angle of $15\degr$.

Summarizing, we find that the simplest explanation of the \fek\ line observed
in BLRGs is a distant torus with $\nh\ga 10^{23}$ cm$^{-2}$ covering a large
solid angle as seen from the nucleus. More complex scenarios are, however,
also compatible with the data.

\subsection{The origin of the Compton reflection}
\label{ss:refl}

A possible explanation of the apparent small solid angle of a Thomson-thick
reflector, $\Omega\ll 2\pi$, is the primary emission originating in the
vicinity of an accretion disk but collimated away from the disk. Collimation is
expected in radio-loud AGNs, e.g., the X\g\ emission of blazars certainly
originates in a jet pointing away from the source plane. For mild collimation,
e.g., one due to a subrelativistic outflow, the observed reduction of Compton
reflection by a factor of a few (with respect to an isotropic source above a
disk) can be easily achieved. In this scenario, the bulk of the \fek\ line does
not originate in the reflection region, but rather in a remote torus (as argued
in Section \ref{ss:line} above). Then we can explain both the approximate
constancy with time of the line flux (from the remote torus) as well as
constant relative weakness of the continuum reflection (from the disk close to
the primary X-ray source). We note that {\it some\/} short time-scale
variability of the line flux is predicted due to the disk-line contribution
(with the expected $\wfe\sim 30$--50 eV at the observed $\Omega$, e.g., GF91).
Also, that disk-line component can be broad due to relativistic effects (e.g.,
Fabian et al.\ 1995).

Alternatively, the cold, reflecting, disk can be truncated at a large radius
(in units of the Schwarzschild radius), due to, e.g., a disk instability. If
the continuum source is located inside the truncation radius, the solid angle
covered by the cold disk would be small. This scenario has been proposed to
explain the weakness of Compton reflection in NGC 4151 (Zdziarski, Johnson \&
Magdziarz 1996).

The observed weak Compton reflection can also be possibly explained in the
absence of a disk by reflection from the torus discussed in Section 7.1 above,
provided $\nh\sim 10^{24}$ cm$^{-2}$ (Krolik, Madau \& \.Zycki 1994;
Ghisellini, Haardt \& Matt 1994). At this $\nh$ there will be already a
signature of spectral hardening above $\sim 10$ keV, but the resulting
reflection spectral component will be substantially weaker than that from a
medium with $\nh\gg 10^{24}$ cm$^{-2}$. This scenario requires some
fine-tuning of $\nh$ and $\Omega_{\rm o}$ but it is in principle possible. Note
that in that case both the \fek\ line and the reflection arise in the same
region and a correlation on long time scales between $\ife$ and $\Omega$ would
be present. The data presented here are insufficient to test for the presence
of such correlation.

Finally, the weakness of Compton reflection could in principle be due to the
dilution of the standard Seyfert-1 component with strong reflection by another
component, e.g., from a comparatively low-luminosity blazar-like jet pointing
close to our line of sight.  This, in principle, would be supported by the
detection of superluminal expansion in the radio images (see Section 4).
However, the X-ray data do not support such a picture. First, the observed
large fluxes of the K$\alpha$ line cannot be directly explained in this
scenario. Second, we have found that the data above $\sim 2$ keV do not allow
the presence of a substantial second continuum component. We have considered a
continuum model consisting of a power law with a free spectral index plus a
(radio-quiet) Seyfert-1--like continuum with fixed $\alpha=0.95$ and
$\Omega/2\pi =0.52$, i.e., the average values obtained by NP94. The line
strength was not tied to either component. We have  found that such a
2-component model provides a fit to the average \ginga\/ spectrum {\it worse\/}
than the single power-law plus reflection model used throughout this paper. The
best fit to the 1-keV normalization of the Seyfert-1 continuum is null, and its
upper limit is 2 per cent of the power-law normalization (at 90 per cent
confidence). Even the upper limit would imply the reflection component much
weaker than the observed $\Omega/2\pi \sim 0.1$--0.3. Thus, Compton reflection
with this $\Omega$ observed in BLRGs cannot be explained as due to dilution of
the Seyfert-1--like X-ray spectrum by another spectral continuum. This
conclusion is supported for 3C 390.3 by observations of a pattern of X-ray
variability implying a single X-ray source (Leighly \& O'Brien 1997).

\subsection{The origin of the continuum and the high-energy break}
\label{ss:break}

As found in the Section 7.2 above, the intrinsic X-ray continua of BLRGs are
unlikely to be composed of two components, e.g., one from the nucleus and one
from a jet. Thus, the X-ray continuum in a BLRG appears to originate
predominantly in a single emission source. This region may be either similar to
that present in radio-quiet Seyfert 1s, or be related to an X\g\ jet, as in
blazars. In any case, the observed continuum emission has to be much more
isotropic than in blazars at least in 3C 390.3, as shown by a Compton
reflection component clearly detected in that BLRG. An important diagnostic for
the nature of the continua can also be provided by the high-energy spectral
breaks observed in the spectra.

We found that the presence of high-energy breaks in the spectra of BLRGs is
established based on: (i) the break in the simultaneous \asca/OSSE spectrum of
3C 120, (ii) the breaks seen in the average spectra of BLRGs observed
repeatedly by both \ginga\/ and OSSE and by both \exosat\/ and OSSE, and (iii)
the average spectra of all BLRGs observed by OSSE being much softer than both
the individual X-ray spectra and the average \ginga\/ and \exosat\/ spectra.
Furthermore, a similar break is clearly seen in Cen A (Kinzer et al.\ 1995), a
narrow-line object considered to differ from BLRGs mostly by orientation (e.g.,
Antonucci 1993). The spectra are seen to break above $\sim 100$ keV, which is
similar to the breaks seen in radio-quiet Seyfert 1s (Z95; Gondek et al.\ 1996;
Zdziarski et al.\ 1996, 1997).

We note that although models with an e-folded power law provide worse spectral
fits to the average X\g\ spectra than the broken power law model (Section 5.3),
there are no available simultaneous X\g\ observations with statistics
sufficient to distinguish between different shapes in soft \g-rays of
individual spectra. A spherical thermal Comptonization model of Poutanen \&
Svensson (1996) including Compton reflection fitted to the average
\ginga/average OSSE spectrum yields the electron temperature of $kT_e\simeq
110$ keV and the optical depth of $\tau\simeq 1.3$ at the best fit, although
the fit is worse than that with the broken power-law model (without reflection,
Table 11), $\Delta \chi^2=+3.4$. A similar thermal Comptonization model fits
well the X-ray to soft \g-ray spectra of radio-quiet Seyfert 1s (e.g.,
Zdziarski et al.\ 1997). That model can also explain the anti-correlation
between the X-ray slope and the ratio between the X-ray to UV fluxes observed
in 3C 120 (Walter \& Courvoisier 1992). Thus, thermal Comptonization remains a
possible explanation of the X\g\ continua of BLRGs.

The relative hardness of the X-ray spectra in BLRGs with respect to radio-quiet
Seyfert 1s can be related in the thermal Comptonization model to the relative
weakness of Compton reflection in BLRGs. This can be explained as follows. In
this model, the soft (UV/soft X-rays) seed photons are repeatedly upscattered
by a hot plasma cloud to the hard X-ray/soft \g-ray range. In the case of
radio-quiet Seyfert, the seed photons are from internal dissipation in an
accretion disk as well as from reprocessing of the incident X\g\ photons (e.g.,
Haardt \& Maraschi 1993). On the other hand, the seed photon field produced
from reprocessing is weak in BLRGs, as inferred from the weakness of the
reflected component. The energy density of the seed photons in the plasma frame
can be further Doppler-reduced if there is a sub-relativistic outflow of the
plasma (see Section \ref{ss:refl} above). From energy balance, the reduced
supply of seed photons leads to reduced cooling and, consequently, to a
hardening of the Comptonized X-ray spectrum (e.g., Poutanen \& Svensson 1996).
Physically, it can be realized, e.g., if the continuum source forms an inner
hot accretion disk thermally Comptonizing cold radiation from the outer disk
and cold clouds within the hot flow, and the reflection of the continuum is
from the outer cold disk (see, e.g., Zdziarski 1998).

On the other hand, the X-ray and soft \g-ray spectra of BLRGs can be due to
emission of nonthermal electrons. This possibility is hinted at by the apparent
power-law shape (with no spectral curvature required) of the average OSSE
spectrum of BLRGs (Section 5.3). Furthermore, the radio emission of BLRGs, and
certainly that arising in the milli-arcsecond region, is nonthermal. The X-ray
spectral slopes of blazars, which emission is certainly nonthermal, are similar
to those found in BLRGs. Also, the blazar spectra universally show a
high-energy spectral break, which energy, however, falls in the MeV range. The
nature of the break is not understood yet (e.g., Sikora et al.\ 1997), but it
is generally agreed upon that their jets show Lorentz factors $\sim 10$; see
Dondi \& Ghisellini (1996). If the plasma in BLRGs propagates with bulk
velocities that are at most trans-relativistic, or, alternatively, they are
viewed at a larger angles to the direction of motion -- the spectral break in
BLRGs could then appear at the appropriately lower energy, $\sim 100$ keV.

Yet another process giving rise to a high-energy spectral break has been
discussed by Skibo, Dermer \& Kinzer (1994). They proposed that the high-energy
break observed in Cen A above $\sim 150$ keV (Kinzer et al.\ 1995) is due to
the emission of a jet (beamed away from our line of sight) being
Compton-scattered by a plasma cloud into our line of sight. The break appears
due to the kinematics of scattering in the Klein-Nishina regime. This model,
however, predicts the break energy at an energy $\ga 1$ MeV at the viewing
angle of $30\degr$, likely to be typical for broad-line objects such as BLRGs.
To achieve a break around 100 keV, as observed for our sample, viewing angles
of $\ga 70\degr$ and relativistic motion of the scattering cloud are necessary
(Fig.\ 1 in Skibo et al.\ 1994). Thus, this model appears to be ruled out for
BLRGs. For Cen A itself, this model has difficulty in explaining the
variability of the soft \g-ray flux on a time scale as short as 12 hours
(Kinzer et al.\ 1995) and the high-energy power law extending to $\sim 10$ MeV
(Steinle et al.\ 1997).

The present difficulty of determining the nature of the high-energy break in
BLRGs is mostly due to the relatively limited sensitivity of the present soft
\g-ray detectors, e.g., OSSE. A progress in resolving this issue is likely
after the launch of {\it INTEGRAL\/} (e.g., Winkler 1996), with its high
sensitivity above 100 keV of the IBIS detector (Ubertini et al.\ 1996).

\section{CONCLUSIONS}
\label{s:con}

The data presented here allow the first systematic comparison of the X\g\
properties of BLRGs with their radio-quiet counterparts, Seyfert 1s. Our main
conclusions are as follows.

1. The intrinsic (i.e., both absorption and reflection-corrected) X-ray
continua of BLRGs are harder on average than those of (radio-quiet) Seyfert 1s.
The \ginga\/ data presented here give the average $\alpha=0.67$ with a
dispersion of 0.18, compared to the average $\alpha=0.95$ with a dispersion of
0.15 in all Seyfert 1s (NP94). Also, the relative strength of Compton-reflected
continuum in BLRGs corresponds to the reflector solid angle of $\ll 2\pi$
whereas the Seyfert-1 data yield significantly larger solid angles, on average
of $\sim (0.5$--$0.7)\times 2\pi$ (NP94). The \asca\/ and \exosat\/ data on
BLRGs presented here are fully compatible with the data from \ginga.

2. The flux in the \fek\ line in BLRGs is in most cases stronger than that
expected from the observed weak Compton reflection. The corresponding average
line equivalent width is $\sim 100$ eV. With a possible exception of 3C 120,
the \asca\/ data are compatible with the \fek\ line being unresolved. The line
flux remains approximately constant in time in spite of the variable continuum
in observations of individual objects spanning several years.

3. The simplest model of the origin of the \fek\ flux is irradiation by the
central X-ray source of a distant torus with $\nh\ga 10^{23}$ cm$^{-2}$. The
torus is away of the direct line of sight in the objects in our sample except
for 3C 445, in which X-ray absorption by such a column density is directly
observed.

4. The data show a high-energy spectral break at $\sim 100$ keV, which
resembles that seen in radio-quiet Seyferts. The break can be explained by
either thermal Comptonization or by nonthermal models, with the latter
marginally favoured by the available data.

5. The main X\g\ spectral properties of BLRGs: the hardness of the X-ray power
law, weak reflection, and a high energy break above 100 keV, are also observed
in the bright narrow-line radio galaxy, Cen A, which would be a BLRG viewed
through the obscuring torus in the framework of the unified AGN model.

\section*{ACKNOWLEDGEMENTS}

This research has been supported in part by the KBN grants 2P03D00614, 
2P03D01209, and 2P03C00511p0(1,4), and NASA grants and contracts.  PRW 
acknowledges a scholarship from T.  Chlebowski, Inc., for undergraduate studies 
in astronomy.  We acknowledge the use of the NASA/GSFC HEASARC archive for the 
\asca\/ and \exosat\/ data.  We thank Julian Krolik and Marek Sikora for 
discussions, and Ian George, Marek Gierli\'nski and Pawe\l\ Magdziarz for their 
valuable assistance with software used in this work.

\bsp

\label{lastpage}

\end{document}